\documentclass[graybox]{svmult}

\usepackage{mathptmx}       
\usepackage{helvet,wrapfig,verbatim}         
\usepackage{courier}        
\usepackage{type1cm}        
\usepackage{color,verbatim} 
\usepackage{makeidx}         
\usepackage{graphicx}        
\usepackage{multicol}        
\usepackage[bottom]{footmisc}
\usepackage{amssymb, amsmath} 
\usepackage{enumerate}
\usepackage{tikz}
\usepackage[outline]{contour}
\contourlength{2pt}

\usepackage{color}

\newcommand{\field}[1]{\mathbb{F}_{#1}}
\newcommand{\indeps}{\mathcal{I}}
\newcommand{\imatroid}{M = (\indeps,E)}
\newcommand{\rmatroid}{M = (\rho,E)}

\newcommand{\linrmatroid}{M[G] = (\rho,E)}
\newcommand{\gammoid}{M(\Gamma,E,T)}
\newcommand{\igammoid}{M(\Gamma,E,T) = (\indeps,E)}
\newcommand{\setmatroid}{M(F_1,\ldots,F_m,E;k;\rho)}
\newcommand{\hcom}{\hbox{, }}
\newcommand{\hand}{\hbox{ and }}

\newcommand{\cl}{\mathrm{cl}}
\newcommand{\e}{\mathrm{cyc}}
\newcommand{\lcode}{[n,k,d] \hbox{-code}}
\newcommand{\cmatroid}{M_C = (\rho_C, E)}
\newcommand{\cflats}{\mathcal{Z}}
\newcommand{\cycles}{\mathcal{U}}

\newcommand{\nkd}{[n,k,d]}
\newcommand{\xnkd}{[n_X,k_X,d_X]}
\newcommand{\xnkdindex}[1]{[n_{#1},k_{#1},d_{#1}]}
\newcommand{\hpar}{(\mathbf{n}, \mathbf{k}, \mathbf{d}, \mathbf{t})}

\newcommand{\colspace}[1]{\mathrm{C}(#1)}
\newcommand{\rowspace}[1]{\mathrm{R}({#1})}
\newcommand{\F}{\mathbb{F}}
\newcommand{\A}{\mathbb{A}}
\newcommand{\gr}{\mathrm{Gr}}
\newcommand{\ee}{\mathrm{e}}
\newcommand{\Z}{\mathbb{Z}}
\newcommand{\FF}{\mathcal{F}}
\newcommand{\UU}{\mathcal{U}}
\newcommand{\ZZ}{\mathcal{Z}}

\newcommand{\apart}{(n,k,d,r,\delta)}
\newcommand{\apar}{(n,k,d,r,\delta,t)}

\newcommand{\ncode}{{(n,k)\hbox{-code}}}

\newcommand{\alp}{A}
\newcommand{\words}{\alp^n}

\newcommand{\cpoly}{P_C = (\rho_C,[n])}

\newcommand{\bblue}{\begin{color}{blue}}
\newcommand{\eblue}{\end{color}}
\newcommand{\bred}{\begin{color}{red}}
\newcommand{\ered}{\end{color}}

\makeindex             

\begin{document}

\title*{Matroid Theory and Storage Codes: Bounds and Constructions}
\author{Ragnar Freij-Hollanti, Camilla Hollanti, and Thomas Westerb\"ack}
\institute{Ragnar Freij-Hollanti, Camilla Hollanti, Thomas Westerb\"ack \at Aalto University, Department of Mathematics and Systems Analysis,
P.O. Box 11100, 
FI-00076 Aalto,
Finland. \email{\{ragnar.freij, camilla.hollanti, thomas.westerback\}@aalto.fi}.
\\
The authors gratefully acknowledge the financial support from the Academy of Finland (grants \#276031 and  \#303819), as well as the support from the COST Action IC1104.
}
%
%
\maketitle

\abstract{Recent research on distributed storage systems (DSSs) has revealed interesting connections between matroid theory and locally repairable codes (LRCs). The goal of this chapter is to introduce the reader to matroids and polymatroids, and illustrate their relation to distributed storage systems. While many of the results are rather technical in nature, effort is made to increase accessibility via simple examples. The chapter embeds all the essential features of LRCs, namely locality, availability, and hierarchy alongside with related generalised Singleton bounds.}

\section{Introduction to Locally Repairable Codes}
In this chapter, we will discuss the theoretical foundations of \emph{locally repairable codes} (LRCs), which were introduced in Chapter 14. While our main interest is in the codes and their applicability for distributed storage systems, significant parts of our machinery comes from matroid theory. We will develop this theory to the extent that is needed for the applications, and leave some additional pointers to interpretations in terms of graphs and projective geometries.

The need for large-scale data storage is continuously increasing. Within the past few years, \emph{distributed storage systems} (DSSs) have revolutionised our traditional ways of storing, securing, and accessing  data. Storage node failure is a frequent obstacle in large-scale DSSs, making repair efficiency an important objective. A bottle-neck for repair efficiency, measured by the notion of \emph{locality} \cite{papailiopoulos12}, is the number of contacted nodes needed for repair. The key objects of study in this paper are \emph{locally repairable codes} (LRCs), which are, informally speaking, storage systems where a  small number of failing nodes can be recovered by boundedly many other (close-by) nodes. Repair-efficient LRCs are already in use for large-scale DSSs used by,  for example, Facebook and Windows Azure Storage \cite{tamo13}.

Another desired attribute, measured by the notion of \emph{availability} \cite{rawat14}, is the property of having multiple alternative ways to repair nodes or access files. This is particularly relevant for nodes containing so-called hot data that is frequently and simultaneously accessed  by  many users. Moreover, as failures are often spatially correlated, it is valuable to have each node repairable at several different \emph{scales}. This means that if a node fails simultaneously with the set of nodes that should normally be used for repairing it, then there still exists a larger set of helper nodes that can be used to recover the lost data. This property is captured by the notion of \emph{hierarchy} \cite{sasidharan15, izs16} in the storage system.

Network coding techniques for large-scale DSSs were considered in \cite{dimakis10}. Since then, a plethora of research on  DSSs with a focus on linear LRCs and various localities has been carried out, see 
\cite{gopalan12, papailiopoulos12, prakash12, silberstein13, tamo14} among many others. Availability for linear LRCs was defined in \cite{rawat14}.  The notion of hierarchical locality was first studied in \cite{sasidharan15}, where bounds for the global minimum distance were also obtained.

Let us denote by $(n,k,d,r,\delta,t)$, respectively, the code length, dimension, global minimum distance, locality, local minimum distance, and availability. Bold-faced parameters $(\mathbf{n}, \mathbf{k}, \mathbf{d}, \mathbf{t})$ will be used in the sequel to refer to hierarchical locality and availability. It was shown in \cite{tamo13} that the $(r,\delta = 2)$-locality of a linear LRC is a matroid invariant. The connection between matroid theory and linear LRCs was examined in more detail in \cite{westerback15}. In addition, the parameters $(n,k,d,r,\delta)$ for linear LRCs were generalised to matroids, and new results for both matroids and linear LRCs were given therein. Even more generally, the  parameters $(n,k,d,r,\delta,t)$ were generalised to polymatroids, and  new results for polymatroids, matroids and both linear and nonlinear LRCs over arbitrary finite alphabets were derived in \cite{polymatroid}. Similar methods can be used to bound parameters of \emph{batch codes} \cite{ZSbatchbound}, as discussed in Chapter~16. For more background on batch codes, see \emph{e.g.} \cite{IKOSbatchfirst, LSbatchorig}. 
Moreover, as certain specific LRCs and batch codes\footnote{To this end, we need to make specific assumptions on the locality and availability of the LRC \cite[Thm. 21]{FVY_PIR}, which also implies restrictions on the query structure of the batch code.} belong to the class of \emph{private information retrieval (PIR)} codes as defined in 
\cite[Def. 4]{FVY_PIR}, the related LRC and batch code bounds also hold for those PIR codes.  See Section 5.3 and Chapter~16 for more discussion.

The main purpose of this chapter is to give an overview of the connection between matroid theory and linear LRCs with availability and hierarchy, using examples for improved clarity of the technical results.  In particular, we are focusing  on how the parameters of a LRC can be analysed using the \emph{lattice of cyclic flats} of an associated matroid, and on a construction derived from matroid theory that provides us with linear LRCs. The matroidal results on LRCs reviewed here are mostly taken from \cite{westerback15, polymatroid,izs16}. 

The rest of this chapter is organised as follows. In Sections 1.1--1.2, we introduce distributed storage systems and how they can be constructed by using linear codes. In particular, we consider locally repairable codes with availability. Section 2 gives a brief introduction to the concepts and features related to matroids relevant to LRCs. In Section 3, we summarise the state-of-the-art generalised Singleton bounds on the code parameters for linear codes, as well as discuss existence of Singleton-optimal linear codes and matroids. Section 4 reviews some explicit (linear) code constructions. In Section 5, we go beyond linear codes and consider polymatroids and related generalised Singleton bounds, which are then valid for all LRCs over any finite alphabet, and also imply bounds for PIR codes when restricted to systematic linear codes. Section 6 concludes the chapter and discusses some open problems. Further results in matroid theory and especially their representability are given in Appendix for the interested reader. 
The following notation will be used throughout the paper:
\medskip

\begin{small}
\begin{tabular}{ll}
$\mathbb{F}$\,: & a field;\\
$\field{q}$\,: & the finite field of prime power size $q$;\\
$E$\,: & a finite set;\\
$G$\,: & a matrix over $\mathbb{F}$ with columns indexed by $E$;\\
$G(X)$\,: & the matrix obtained from $G$ by restricting to the\\
      & columns indexed by $X$, where $X \subseteq E$;\\
$\colspace{G}$\,: & the vector space generated by the columns of $G$;\\
$\rowspace{G}$\,: & the vector space generated by the rows  of $G$;\\
$C$\,: & linear code $C = \rowspace{G}$ over $\mathbb{F}$ generated by $G$; \\
$C_X$\,: & the \emph{punctured} code of $C$ on $X$, {\em i.e.},\\
    & $C_X = \rowspace{G(X)}$,  where $X \subseteq E$;\\
$2^E$\,: & the collection of all subsets of a finite set $E$;\\ 
$\lbrack j \rbrack$\,: & the set $\{1,2,\ldots,j\}$ for a positive integer $j$;\\
$n,k,d,r,\delta,t,h$\,: & code length, dimension, minimum distance, locality, \\ 
& failure tolerance, availability, hierarchy, respectively;\\
$[n,k,d]\,, (n,k,d)$\,: & parameters of a linear/general code, respectively;\\
$(n,k,\ldots)_i$\,: & parameter values when we consider information \\ 
&symbols, \emph{e.g.}, information symbol locality;\\ 
$(n,k,\ldots)_a$\,: & parameter values when we consider all code \\ 
&symbols, \emph{e.g.}, all symbol locality;\\ 
$(n,k,\ldots)_s$\,: & parameter values when we consider systematic code \\ 
&symbols, \emph{e.g.}, systematic symbol locality;\\ 
$(\mathbf{n}, \mathbf{k}, \mathbf{d}, \mathbf{t})$\,: & parameter values for different hierarchy levels.
\end{tabular}
\end{small}


\begin{remark}
Here, $d$ denotes the minimum (Hamming) distance of the code, rather than the number of nodes that have to be contacted for repair, as is commonplace in the theory of regenerating codes. In Chapter 14, the minimum distance of the code was denoted by $d_H$. 

The motivation to study punctured codes arises from hierarchical locality; the locality parameters at the different hierarchy levels correspond to the global parameters of the related punctured codes. The puncturing operation on codes corresponds to the so-called restriction (or deletion) operation on matroids.

We also point out that $G(E) = G$ and $C_E = C$. We will often index a matrix $G$ by $[n]$, where $n$ is the number of columns in $G$. 
\end{remark}

\subsection{Distributed Storage Systems from Linear Codes}
A linear code $C$ can be used to obtain a DSS, where every coordinate in $C$ represents a storage node in the DSS, and every point in $C$ represents a stored data item. While one often assumes that the data items are field elements in their own right, no such assumption is necessary. However, if $C$ is a code over the field $\F$ and the data items are elements in an alphabet $\A$, then we must be able to form the linear combinations $f_1 a_1 + f_2 a_2$ for $f_1, f_2$ in $\F$ and $a_1, a_2$ in $\A$. Moreover, if we know the scalar $f$, we must be able to read off $a$ from $f a$. This is achieved if $\A\cong\F^\alpha$ is a vector space over $\F$, wherefore we must have $|\A|\geq |\F|$. 
Thus, the length of the data items must be at least the number of symbols needed to represent a field element. In particular if the data items are measured in, \emph{e.g.}, kilobytes, then we are restricted to work over fields of size not larger than about $2^{8000}$. Beside this strict upper bound on the field size, the complexity of operations also makes small field sizes --- ideally even binary fields --- naturally desirable.
\begin{example}\label{ex:dss_code}
Let $C$ be the linear code generated by the following matrix $G$ over $\field{3}$:  
$$
\begin{small}
G=
\begin{tabular}{ |c|c|c|c|c|c|c|c|c| }
\multicolumn{1}{c}{1}&
\multicolumn{1}{c}{2}&
\multicolumn{1}{c}{3}&
\multicolumn{1}{c}{4}&
\multicolumn{1}{c}{5}&
\multicolumn{1}{c}{6}&
\multicolumn{1}{c}{7}&
\multicolumn{1}{c}{8}&
\multicolumn{1}{c}{9}\\
\hline
1&0&0&0&1&1&1&1&1\\
\hline
0&1&0&0&1&0&1&2&2\\
\hline
0&0&1&0&0&1&1&0&0\\
\hline
0&0&0&1&0&0&0&1&2\\
\hline
\end{tabular}
\end{small}
$$
Then, $C$ corresponds to a $9$ node storage system, storing four files $(a,b,c,d)$, each of which is an element in $\field{3}^\alpha$. In this system, node $1$ stores $a$, node $5$ stores $a+b$, node $9$ stores $a+2b+2d$, and so on. 
\end{example}

 Two very basic properties of any DSS are that every node can be repaired by some other nodes and that every node contains some information\footnote{We remark that if one takes into account queueing theoretic aspects, then data allocation may become less trivial (some nodes may be empty). Such aspects are discussed, especially in a wireless setting, in Chapters 12 and 13. However, these considerations are out of the scope of this chapter.}. We therefore give the following definition.

\begin{definition} \label{def:LRC}
A linear $\nkd$-code $C$ over a field is a \emph{non-degenerate storage code} if $d \geq 2$ and there is no zero column in a generator matrix of $C$.
\end{definition}

The first example of a storage code, and the motivating example behind the notion of locality, is the notion of a \emph{maximum distance separable (MDS) code}. It has several different definitions in the literature, here we list a few of them.

\begin{definition}
The following properties are equivalent for linear $[n,k,d]$ storage codes:
\begin{enumerate}[(i)]
\item $n=k+d-1$.
\item The stored data can be retrieved from any $k$ nodes in the storage system.
\item To repair any single erased node in the storage system, one needs to contact $k$ other nodes.
\end{enumerate}
A code that satisfies one (and therefore all) of the above properties is called an $MDS$ code. \end{definition}

By the Singleton bound, $n\geq k+d-1$ holds for any storage code, so by property (i), MDS codes are ``optimal'' in the sense that they have minimal length for given storage capacity and error tolerance. However, (iii) is clearly an unfavourable property in terms of erasure correction. This is the motivation behind constructing codes with small $n-k-d$, where individual node failures can still be corrected ``locally''.

\subsection{Linear Locally Repairable Codes with Availability}

The very broad class of linear LRCs will be defined next. It is worth noting that, contrary to what the terminology would suggest, a LRC is not a novel kind of code, but rather the ``locality parameters'' $(r,\delta)$ can be defined for any code. What we call a LRC is then only a code that is specifically designed with the parameters $(r,\delta)$ in mind. While the locality parameters can be understood directly in terms of the storage system, it is more instructive from a coding theoretic point of view to understand them via punctured codes. Then, the punctured codes will correspond exactly to the restrictions to  ``locality sets'', which can be used to locally repair a small number of node failures within the locality set.

\begin{definition}\label{def:param_codes} Let $G$ be a matrix over $\mathbb{F}$ indexed by $E$ and $C$ the linear code generated by $G$. Then, for $X \subseteq E$, $C_X$ is a linear $\xnkd$-code where
$$ 
\begin{array}{l}
n_X = |X|,\\
k_X = \mathrm{rank}(G(X)),\\
d_X = \min \{|Y| : Y \subseteq X \hand k_{X \setminus Y} < k_X\}. 
\end{array}
$$
\end{definition}

Alternatively, one can define the minimum distance $d_X$ as the smallest support of a non-zero codeword in $C_X=\rowspace{G(X)}$. We use Definition~\ref{def:param_codes}, as it has the advantage of not depending on the linearity of the code.

\begin{example} \label{ex:nX_kX_dX-matrix} 
Consider the storage code $C$ from Example~\ref{ex:dss_code}. Let $Y_1 = \{1,2,3,5,6,7\}$, $X_1 = \{1,2,5\}$ and $X_2 = \{2,6,7\}$.
Then $C_{Y_1}$, $C_{X_1}$ and $C_{X_2}$ are storage codes with 
$$
\begin{array}{lcl}
\xnkdindex{Y_1}&=&[6,3,3]\,,\\
\xnkdindex{X_1}&=&[3,2,2]\,,\\
\xnkdindex{X_2}&=&[3,2,2]\,.
\end{array}
$$
\end{example}

The parameter $d_X$ is the minimum (Hamming) distance of $C_X$. We say that $C$ is an $[n,k,d]$-code with $[n,k,d] = [n_E,k_E, d_E]$.

We choose the following definition for \emph{general} $(n,k,d,r,\delta,t)$-LRCs (\emph{i.e.}, both linear and nonlinear), which we will compare  to known results for \emph{linear} LRCs.    

\begin{definition}
An \emph{$(n,k,d)$-code} $C$ over $A$ is a nonempty subset $C$ of $A^n$, where $A$ is a finite set of size $s$, $k = \log_s(|C|)$, and $d$ the minimum (Hamming) distance of the code.  For $X = \{i_1,\ldots,i_m\} \subseteq E$, the puncturing $C_X$ is defined as $$C_X  = \{(c_{i_1},\ldots,c_{i_m}): \boldsymbol{c} \in C \}.$$
The code $C$ is \emph{non-degenerate}, if $d \geq 2$ and $|C_{\{i\}}| > 1$ for all coordinates $i \in [n]$.  
\end{definition}

\begin{definition} \label{def:LRC_local_avail}
A \emph{locally repairable code} over $A$ is a non-degenerate $(n,k,d)$-code $C$. A coordinate $x \in [n]$ of $C$ has \emph{locality} $(r,\delta)$ and \emph{availability} $t$ if there are $t$ subsets $R_1,\ldots,R_t$ of $[n]$, called repair sets of $x$, such that for $i,j \in [t]$
$$
\begin{array}{rl}
(i) & x \in R_i,\\
(ii) & |R_i| \leq r + \delta - 1,\\
(iii) & d(C_{R_i}) \geq \delta,\\
(iv) & i \neq j \quad \Rightarrow \quad R_i \cap R_j = \{x\}.
\end{array}
$$  
\end{definition}

If every element $x\in X\subseteq E$ has availability with parameters $\apar$ in $C$, then we say that the set $X$ has $\apar$-availability in $C$. We will often talk about codes with $\apart$-locality, by which we mean a code that has $(n,k,d,r,\delta, 1)$-availability, so that symbols are not required to be included in more than one repair set. If the other parameters are clear from the context, we may shortly say that $X$ has ``locality $(r,\delta)$'' or ``availability $t$'', along the lines of the above definition.

An \emph{information set} of a linear $\nkd$-code $C$ is defined as a set $X \subseteq E$ such that $k_X = |X| = k$. Hence, $X$ is an information set of $C$ if and only if there is a generator matrix $G$ of $C$ such that $G(X)$ equals the identity matrix, {\em i.e.}, $C$ is systematic in the coordinate positions indexed by $X$ when generated by $G$. In terms of storage systems, this means that the nodes in $X$ together store all the information of the DSS.

\begin{example}\label{ex:info1}
Two examples of an information set of the linear code $C$ generated by $G$ in Example \ref{ex:dss_code} are $\{1,2,3,4\}$ and $\{1,2,6,8\}$. 
\end{example}

More formally we define: 
\begin{definition}
Let $C$ be an $(n,k,d)$-code and $X$ a subset of $[n]$. Then $X$ is an \emph{information set} of $C$ if $\log_s(|C_X|) = k$ and $\log_s(C_Y)<k$ for all $Y \subsetneq X$. Further, $X$ is \emph{systematic} if $k$ is an integer, $|X| = k$ and $C_X = A^k$. Also, $X$ is an \emph{all-coordinate set} if $X = [n]$. 
\end{definition}

\begin{definition} \label{def:s_i_a_LRC}
A \emph{\underline{s}ystematic-symbol, \underline{i}nformation-symbol}, and \emph{\underline{a}ll-symbol} LRC, respectively, is an $(n,k,d)$-LRC with a systematic, information set and all-coordinate set $X$, such that every coordinate in $X$ has locality $(r,\delta)$ and availability $t$. These are denoted by 
$$
(n,k,d,r,\delta,t)_s\mathrm{-LRC},\ (n,k,d,r,\delta,t)_i\mathrm{-LRC}, \mathrm{\ and\ } (n,k,d,r,\delta,t)_a\mathrm{-LRC},
$$ 
respectively. Further, when availability is not considered ($t = 1$), we get natural notions of  $(n,k,d,r,\delta)_s$, $(n,k,d,r,\delta)_i$, and $(n,k,d,r,\delta)_a$-LRCs. 
\end{definition}

\section{Introduction to Matroids}
\label{sec:matroids}

Matroids were first introduced by Whitney in 1935, to capture and generalise the notion of linear dependence in purely combinatorial terms~\cite{whitney35}. Indeed, the combinatorial setting is general enough to also capture many other notions of dependence occurring in mathematics, such as cycles or incidences in a graph, non-transversality of algebraic varieties, or algebraic dependence of field extensions. Although the original motivation comes from linear algebra, we will see that a lot of matroid terminology comes from graph theory and projective geometry. More details about these aspects of matroid theory are relegated to the appendix.

\subsection{Definitions}
We begin by presenting two equivalent definitions of a matroid.

\begin{definition}[Rank function]\label{rankdef} 
A \emph{(finite) matroid} $\rmatroid$ is a finite set $E$ together with a \emph{rank function} $\rho:2^E \rightarrow \mathbb{Z}$ such that for all subsets $X,Y \subseteq E$
$$
\begin{array}{rl} 
(R.1) & 0 \leq \rho(X) \leq |X|,\\
(R.2) & X \subseteq Y \quad \Rightarrow \quad \rho(X) \leq \rho(Y),\\
(R.3) & \rho(X) + \rho(Y) \geq \rho(X \cup Y) + \rho(X \cap Y). 
\end{array}
$$ 
\end{definition}

An alternative but equivalent definition of  a matroid is the following.

\begin{definition}[Independent sets]\label{indepdef} 
A \emph{(finite) matroid} $\imatroid$ is a finite set $E$ and a collection of subsets $\indeps \subseteq 2^E$ such that
$$
\begin{array}{rl} 
(I.1) & \emptyset \in \indeps,\\
(I.2) & Y \in \indeps \hcom X \subseteq Y \Rightarrow X \in \indeps,\\
(I.3) & \hbox{For all pairs }X,Y \in \indeps \hbox{ with } |X| < |Y|,\hbox{ there exists } \\ & y \in Y \setminus X \hbox{ such that } X \cup \{y\} \in \indeps. 
\end{array}
$$ 
The subsets in $\indeps$ are the \emph{independent sets} of the matroid. 
\end{definition}

The rank function $\rho$ and the independents sets $\indeps$ of a matroid on a ground set $E$ are linked as follows: For $X \subseteq E$,
$$
\rho(X) = \max \{|Y| : Y \subseteq X \hand Y \in \indeps\},$$ and 
$X \in \indeps$ if and only if $\rho(X) = |X|$. It is an easy (but not trivial) exercise to show that the two definitions are equivalent under this correspondence. Another frequently used definition is in terms of the set of \emph{bases} for a matroid, which are the maximal independent sets.  Further, we will also use the \emph{nullity function} $\eta: 2^E \rightarrow \mathbb{Z}$, where $\eta(X) = |X| - \rho(X) $ for $X \subseteq E$.

Any matrix $G$ over a field $\F$ generates a matroid $M_G=(\rho,E)$, where $E$ is the set of columns of $G$, and $\rho(I)$ is the rank over $\F$ of the induced matrix $G(I)$ for $I\subseteq E$. Consequently, $I\in\indeps$ precisely when $I$ is a linearly independent set of vectors. It is straightforward to check that this is a matroid according to Definition \ref{rankdef}. As elementary row operations preserve the row space $R(G(I))$ for all $I\subseteq E$, it follows that row equivalent matrices generate the same matroid.

Two matroids $M_1 = (\rho_1, E_1)$ and $M_2 = (\rho_2,E_2)$ are \emph{isomorphic} if there exists a bijection $\psi: E_1 \rightarrow E_2$ such that $\rho_2(\psi(X)) = \rho_1(X)$ for all subsets $X \subseteq E_1$.

\begin{definition}A matroid that is isomorphic to $M_G$ for some matrix $G$ over $\F$ is said to be \emph{representable} over $\F$. We also say that such a matroid is \emph{$\F$-linear}.\end{definition}

Two trivial matroids are the zero matroid where $\rho(X)=0$ for each set $X\subseteq E$, and the one where $\rho(X)=|X|$ for all $X\subseteq E$. These correspond to all-zeros matrices and invertible $n\times n$-matrices respectively. The first non-trivial example of a matroid is the following:

\begin{definition}
The \emph{uniform matroid} $U_n^k=(\rho, [n])$, where $[n]=\{1,2,\cdots , n\}$, is given by the rank function $\rho(X)=\min\{|X|, k\}$ for $X\subseteq [n]$.  
\end{definition}

The following straightforward observation gives yet another characterisation of MDS codes.

\begin{proposition}
$G$ is the generator matrix of an $[n,k,n-k+1]$-MDS code if and only if $M_G$ is the uniform matroid $U_n^k$.
\end{proposition}

\subsection{Matroid Operations}
For explicit constructions of matroids, as well as for analysing their structure, a few elementary operations are useful. Here, we will define these in terms of the rank function,
but observe that they can equally well be formulated in terms of independent sets. The effect of these operations on the representability of the matroid is discussed in Appendix. In addition to the operations listed here, two other very important matroid operations are dualisation and contraction. As these are not explicitly used here to understand locally repairable codes, we leave their definition to Appendix.

\begin{definition}
The \emph{direct sum} of two matroids $M=(\rho_M,E_M)$ and $N=(\rho_N,E_N)$ is $$M\oplus N =(\tau, E_M\sqcup E_N),$$ where $\sqcup$ denotes the disjoint union, and $\tau:2^{E_M\sqcup E_N}\to \Z$ is defined by $\tau(X)=\rho_M(X\cap E_M)+\rho(X\cap E_N)$.
\end{definition}
Thus, all dependent sets from $M$ and $N$ remain dependent in $M\oplus N$, whereas there is no dependence between elements in $M$ and elements in N. If $M$ and $N$ are graphical matroids\footnote{See Appendix for the definition of graphical matroids}, then $M\oplus N$ is graphical, and obtained from the disjoint union of the graphs associated to $M$ and $N$. 

\begin{definition} The \emph{restriction} of $\rmatroid$ to a subset $X\subseteq E$ is the matroid $M_{|X} = (\rho_{|X},X)$, where
\begin{equation} \label{eq:definition_matroid_restriction}
\rho_{|X}(Y) = \rho(Y) \hbox{, for } Y \subseteq X.
\end{equation} 
\end{definition}

Obviously, for any matroid $M$ with underlying set $E$, we have $M_{|E} = M$. The restriction operation is also often referred to as \emph{deletion} of $E\setminus X$, especially if $E\setminus X$ is a singleton. Given a matrix $G$ that represents $M$, the submatrix $G(X)$ represents $M_{|X}$.

\begin{definition}
The truncation of a matroid $\rmatroid$ at rank $k\leq \rho(E)$ is $M_k=(\rho',E)$, where $\rho'(X)=\min\{\rho(X),k\}$.
\end{definition}
Geometrically, the truncation of a matroid corresponds to projecting a point configuration onto a \emph{generic} $k$-dimensional space. However, this does not imply that truncations of $\F$-linear matroids are necessarily $\F$-linear, as it may be the case that there exists no $k$-space that is in general position relative to the given point configuration. However, it is easy to see that $M_k$ is always representable over some field extension of $\F$. In fact, via a probabilistic construction, one sees that the field extension can be chosen to have size at most $q\binom{n}{k}$~\cite{jurrius12}.

The \emph{relaxation} is the elementary operation that is most difficult to describe in terms of rank functions. It is designed to destroy representability of matroids, and corresponds to selecting a hyperplane in the point configuration, and perturbing it so that its points are no longer coplanar. To prepare for the definition, we say that a \emph{circuit} is a dependent set, all of whose subsets are independent. For any nonuniform matroid $\rmatroid$, there are circuits of rank $\rho(E)-1$. This is seen by taking any dependent set of rank $\rho(E)-1$, and deleting elements successively in such a way that the rank does not decrease. We mention that a matroid that has no circuits of rank $<\rho(E)-1$ is called a \emph{paving matroid}. It is conjectured that asymptotically (in the size) almost all matroids are paving~\cite{mayhew11}. Recent research shows that this is true at least on a ``logarithmic scale''~\cite{pendavingh15}. 

\begin{definition}
Let $\imatroid$ be a matroid with rank function $\rho$, and let $C$ be a circuit of rank $\rho(E)-1$. The \emph{relaxation} of $M$ at $C$ is the matroid $(\mathcal{I}\cup\{C\}, E)$.
\end{definition}

\begin{figure}[hbt]
\sidecaption
\includegraphics[scale=.65]{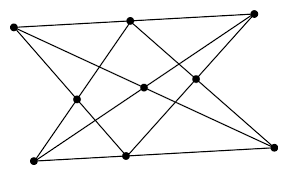}
\caption{The non-Pappus matroid, which is not representable over any field. If there were a line between the three points in the middle, then the figure would illustrate the matroid $M_G$ in Example~\ref{ex:pappus}. Relaxation of the circuit $\{4,5,6\}$ corresponds to deletion of this line in the figure.}
\label{fig:pappus}       
\end{figure}

\begin{example}\label{ex:pappus}The first example of a matroid constructed by relaxation is the \emph{non-Pappus matroid} of Figure~\ref{fig:pappus}. This is constructed by relaxing the circuit $\{4,5,6\}$ from the representable matroid $M_G$, where 
\[G=
\begin{tabular}{ |c|c|c|c|c|c|c|c|c| }
\multicolumn{1}{c}{1}&
\multicolumn{1}{c}{2}&
\multicolumn{1}{c}{3}&
\multicolumn{1}{c}{4}&
\multicolumn{1}{c}{5}&
\multicolumn{1}{c}{6}&
\multicolumn{1}{c}{7}&
\multicolumn{1}{c}{8}&
\multicolumn{1}{c}{9}\\
\hline
1&0&-1&1/2&0&-1/2&1&0&-1\\
\hline
1&1&1&0&0&0&-1&-1&-1\\
\hline
1&1&1&1&1&1&1&1&1\\
\hline
\end{tabular}\] and can be defined over any field of odd (or zero) characteristic, other than $\F_3$.\end{example}

\subsection{Matroid Invariants of Codes}
 There is a straightforward connection between linear codes and matroids. Indeed, let $C$ be a linear code generated by a matrix $G$. Then $C$ is associated with the matroid $M_G = (\rho, E)$. As two different generator matrices of $C$ have the same row space, they will generate the same matroid. Therefore, without any inconsistency, we denote the associated linear matroid of $C$ by $\cmatroid$. In general, there are many different codes $C\neq C'$ with the same matroid structure $M_C = M_{C'}$.  In the appendix, we will see how this phenomenon can be interpreted as a stratification of the Grassmannian over a finite field. 
 
A property of linear codes that depends only on the matroid structure of the code is called \emph{matroid invariant}. For example, the collection of information sets and the parameters $\nkd$ of a code are matroid invariant properties. This is the content of the following easy proposition.

\begin{proposition}
Let $C$ be a linear $\nkd$-code and $X \subseteq E$. Then for $\cmatroid$,
$$
\begin{array}{rl}
(i) & n_X = |X|,\\
(ii) & k_X = \rho_C(X),\\
(iii) & d_X = \min \{|Y| : Y \subseteq X \hcom \rho_C(X \setminus Y) < \rho_C(X)\},\\
(iv) & X \hbox{ is an information set of $C$} \iff \\
& X \hbox{ is a basis  of $M_C$} \iff \rho_C(X) = |X| = k.
\end{array}
$$
\end{proposition}

In addition to the parameters $\nkd$ of a linear code $C$, we are also interested in the length, rank and minimum distance of the punctured codes, since these correspond to the locality parameters at the different hierarchy levels, which we will discuss in more detail in Section \ref{sec:beyond_linear_storage}.

 A punctured code can be analysed using matroid restrictions, since $M_{C|X} = M_C|X$ for every coordinate subset $X$. Thus, the parameters $\xnkd$ of $C_X$ are also matroid invariant properties for $C$.

\begin{example}
Let $C$ denote the $\nkd$-code generated by the matrix $G$ given in Example \ref{ex:dss_code}. Then $\nkd = [9,4,3]$, where the value of $d$ arises from the fact that $\rho_C([9] \setminus \{i,j\})= 4$ for $i,j = 1,2,\ldots,9$, and $\rho_C([9]\setminus\{4,8,9\}) = 3$. Two information sets of $C$ are $\{1,2,3,4\}$ and $\{1,2,6,8\}$, as we already saw before in Example \ref{ex:info1}. 
\end{example}

It is rather easy to see that two different linear codes can have the same associated matroid. As a consequence, not every property of a linear code is matroid invariant. An example of a code invariant that is not matroid invariant is the covering radius \cite{britz05, skorobogatov92}. Indeed, an $[n,k,d]$-MDS code, \emph{i.e.}, a realisation of the uniform matroid $U_n^k$, generically has covering radius $d-1=n-k$, yet there exist MDS codes with lower covering radii. An explicit example is given in~\cite{britz05}.

\subsection{The Lattice of Cyclic Flats}
One matroid invariant that has singled out as essential for describing the repairability of storage codes is the lattice of cyclic flats. To define this, remember that $X\subseteq E$ is a \emph{circuit} in $\imatroid$ if $X$ is dependent, but all proper subsets of $X$ are independent. A \emph{cyclic} set is a (possibly empty) union of circuits. Equivalently, $X$ is \emph{cyclic} if for every $x \in X$ 
$$
\rho(X \setminus \{x\}) = \rho(X). 
$$
Let us define the operation $\mathrm{cyc}:2^E\to 2^E$ by
$$
\mathrm{cyc}(X) = \{x \in X : \rho(X \setminus \{x\}) = \rho(X)\}.
$$ Then $X$ is cyclic if and only if $\mathrm{cyc}(X)=X$. We refer to $\e$ as the \emph{cyclic core} operator.
 
Dually, we define the \emph{closure} of $X$ to be 
$$
\cl(X) =  \{y \in E : \rho(X \cup \{y\}) = \rho(X)\}, $$        
and notice that $X\subseteq \cl(X)$ by definition. We say that $X$ is a \emph{flat} if $\cl(X) = X$. Therefore, $X$ is a \emph{cyclic flat} if 
$$
\rho(X \setminus \{x\}) = \rho(X) \hand \rho(X \cup \{y\}) > \rho(X)
$$ 
for all $x \in X$ and $y \in E \setminus X$. The set of flats, cyclic sets, and cyclic flats of $M$ are denoted by $\FF(M)$, $\UU(M)$, and $\ZZ(M)$, respectively. 

It is not entirely obvious that the set of cyclic flats is nonempty. However, it follows from the matroid axioms that the closure operator $\cl$ preserves cyclicity, and that the cyclic core operator $\e$ preserves flatness. Thus we can consider $\cl$ and $\e$ as maps \[\cl: \left\{\begin{split} 2^E & \to\FF(M)\\ \UU(M)&\to\ZZ(M)\end{split}\right.\,,\] and $$\e: \left\{\begin{split} 2^E & \to\UU(M)\\ \FF(M)&\to\ZZ(M)\end{split}\right.\,.$$ In particular, for any set $X\subseteq E$, we have $\e\circ\cl(X)\in\ZZ(M)$ and $\cl\circ\e(X)\in\ZZ(M)$.

 
Let $\linrmatroid$ be a linear matroid, generated by $G$. Then $X \subseteq E$ is a cyclic flat if and only if the following two conditions are satisfied
$$
\begin{array}{rl}
(i) & \colspace{G(X)} \cap \colspace{G(E \setminus X)} = \mathbf{0}\\
(ii) & x \in X \Rightarrow \colspace{G(X \setminus \{x\})} = \colspace{G(X)}.
\end{array}
$$ 
In terms of storage codes, a cyclic flat is thus a set $X \subseteq E$ of storage nodes such that every node in  $X$ can be repaired by the other nodes in $X$, whereas no node outside $X$ can be repaired by $X$. This observation shows the relevance of cyclic flats for storage applications. The strength of using $\ZZ(M)$ as a theoretical tool comes from its additional \emph{lattice structure}, which we will discuss next.  


A collection of sets $\mathcal{P} \subseteq 2^E$ ordered by inclusion defines a partially ordered set (poset) $(\mathcal{P}, \subseteq)$. Let $X$ and $Y$ denote two elements of $\mathcal{P}$. $Z$ is the \emph{join} $X \vee Y$ if it is the unique maximal element in $\{W \in \mathcal{P} \,:\, X \subseteq W \hcom Y \subseteq W\}$. Dually, $Z$ is the \emph{meet} $X \wedge Y$ if it is the unique minimal element in $\{W \in \mathcal{P}\,:\, X \supseteq W \hcom Y \supseteq W\}$. 

A pair of elements in an arbitrary poset does not need to have a join or a meet. If $(\mathcal{P}, \subseteq)$ is a poset such that every pair of elements in $\mathcal{P}$ has a join and a meet, then $\mathcal{P}$ is called a \emph{lattice}. The bottom and top elements of a finite lattice $(\mathcal{P}, \subseteq)$ always exist, and are denoted by $1_{\mathcal{P}} = \bigvee_{X \in \mathcal{P}} X$ and $0_{\mathcal{P}}= \bigwedge_{X \in \mathcal{P}} X$, respectively.

Two basic properties of cyclic flats of a matroid are given in the following proposition.
\begin{proposition} [\cite{bonin08}] \label{prop:basic-Z}
Let $M = (\rho,E)$ be a matroid and $\cflats$ the collection of cyclic flats of $M$. Then, 
\begin{enumerate}
\item[(i)] $\rho(X) = \min \{\rho(F) + | X \setminus F | : F \in \mathcal{Z}\}$, for $X \subseteq E$,
\item[(ii)] $(\mathcal{Z}, \subseteq)$ is a lattice, $X \vee Y = \mathrm{cl}(X \cup Y)$ and \\
 $X \wedge Y = \mathrm{cyc}( X \cap Y) $ for $X,Y \in \mathcal{Z}$\,.
\end{enumerate}
\end{proposition}

Proposition \ref{prop:basic-Z} (i) shows that a matroid is uniquely  determined by its cyclic flats and their ranks. 

\begin{example} \label{ex:Z}
Let $\cmatroid$ be the matroid associated to the linear code $C$ generated by the matrix $G$ given in Example \ref{ex:dss_code}. The lattice of cyclic flats $(\cflats,\subseteq)$ of $M_C$ is given in the figure below, where the cyclic flat and its rank are given at each node.
$$
\begin{tikzpicture}
    \node[shape=circle,draw=black] (00) at (0,0) {$0_\cflats$};  
    \node[shape=circle,draw=black] (11) at (-3.6,2) {$X_1$}; 
    \node[shape=circle,draw=black] (12) at (-1.8,2) {$X_2$}; 
    \node[shape=circle,draw=black] (13) at (0,2) {$X_3$}; 
    \node[shape=circle,draw=black] (14) at (1.8,2) {$X_4$}; 
    \node[shape=circle,draw=black] (15) at (3.6,2) {$X_5$}; 
    \node[shape=circle,draw=black] (21) at (-2,4) {$Y_1$}; 
    \node[shape=circle,draw=black] (22) at (2,4) {$Y_2$}; 
    \node[shape=circle,draw=black] (31) at (0,5) {$1_\cflats$};    
    
     \path [ultra thin] (00) edge (11);
     \path [ultra thin] (00) edge (12);
     \path [ultra thin] (00) edge (13);
     \path [ultra thin] (00) edge (14);
     \path [ultra thin] (00) edge (15);
     \path [ultra thin] (11) edge (21);
     \path [ultra thin] (12) edge (21);
     \path [ultra thin] (13) edge (21);
     \path [ultra thin] (14) edge (21);
     \path [ultra thin] (14) edge (22);
     \path [ultra thin] (15) edge (22);
     \path [ultra thin] (21) edge (31);
     \path [ultra thin] (22) edge (31);
    
    \node  [fill=white,rounded corners=2pt,inner sep=1pt] (t00) at (0,-0.7) { \small{$(\emptyset,0)$ }}; 
    \node  [fill=white,rounded corners=2pt,inner sep=1pt] (t11) at (-3.6,2.6) {\small{ $(\{1,2,5\},2)$ }}; 
    \node [fill=white,rounded corners=2pt,inner sep=1pt] (t12) at (-1.8,1.4) {\small{ $(\{2,6,7\},2)$ }}; 
    \node [fill=white,rounded corners=2pt,inner sep=1pt] (t13) at (-0,2.6) { \small{$(\{3,5,7\},2)$ }}; 
    \node [fill=white,rounded corners=2pt,inner sep=1pt] (t14) at (1.8,1.4) { \small{$(\{1,3,6\},2)$ }}; 
    \node  [fill=white,rounded corners=2pt,inner sep=1pt] (t15) at (3.6,2.6) { \small{$(\{4,8,9\},2)$ }}; 
    \node [fill=white,rounded corners=2pt,inner sep=1pt] (t21) at (-2,4.6) {\small{ $(\{1,2,3,5,6,7\},3)$ }}; 
    \node [fill=white,rounded corners=2pt,inner sep=1pt] (t22) at (2,4.6) { \small{$(\{1,2,4,5,8,9\},3)$ }}; 
    \node [fill=white,rounded corners=2pt,inner sep=1pt] (t31) at (0,5.7) { \small{$([9],4)$ }}; 

\end{tikzpicture}
$$
\end{example}

An axiom scheme for matroids via cyclic flats and their ranks was independently given in~\cite{bonin08} and \cite{sims80}.
This gives a compact way to construct matroids with prescribed local parameters, which we have exploited in~\cite{westerback15}.

\begin{theorem} [see \cite{bonin08} Th. 3.2 and \cite{sims80}] \label{th:Z-axiom}
Let $\mathcal{Z} \subseteq 2^E$ and let $\rho$ be a function $\rho: \mathcal{Z} \rightarrow \mathbb{Z}$. There is a matroid $M$ on $E$ for which $\mathcal{Z}$ is the set of cyclic flats and $\rho$ is the rank function restricted to the sets in $\mathcal{Z}$, if and only if
$$
\begin{array}{rl}
(Z0) & \mathcal{Z} \hbox{ is a lattice under inclusion},\\
(Z1) & \rho(0_{\mathcal{Z}}) = 0,\\
(Z2) & X,Y \in \mathcal{Z} \hbox{ and } X \subsetneq Y \Rightarrow \\
     & 0 < \rho(Y) - \rho(X) < | Y | - | X |,\\
(Z3) & X,Y \in \mathcal{Z} \Rightarrow \rho(X) + \rho(Y) \geq \\
     & \rho(X \vee Y) + \rho(X \wedge Y) + | (X \cap Y) \setminus (X \wedge Y)  |.
\end{array}
$$
\end{theorem}

For a linear $\nkd$-code $C$ with $\cmatroid$ and $\cflats = \cflats(M_C)$, and for a coordinate $x$, we have
$$
\begin{array}{rl}
(i) & d \geq 2 \iff 1_{\cflats} = E,\\
(ii) & C_{\{x\}} \neq \{0_\mathbb{F}\} \hbox{ for every } x \in E \iff 0_{\cflats} = \emptyset.
\end{array}
$$
Hence, by Definition \ref{def:LRC}, we can describe non-degeneracy in terms of the lattice of cyclic flats, as follows.

\begin{proposition}
Let $C$ be a linear $\lcode$ and $\cflats$ denote the collection of cyclic flats of the matroid $\cmatroid$. Then $C$ is a non-degenerate storage code if and only if $0_\cflats = \emptyset$ and $1_\cflats = E$.
\end{proposition}

\begin{proposition}
Let $C$ be a non-degenerate storage code and $\cmatroid$. Then, for $X \subseteq E$,
$C_X$ is a non-degenerate storage code if and only if $X$ is a cyclic set of $M_C$.
\end{proposition} 

As cyclic sets correspond to non-degenerate subcodes, and hence to systems where every symbol is stored with redundancy, we will use these as our ``repair sets''. Therefore, we want to determine from the lattice of cyclic flats, whether a set is cyclic or not, which we achieve through the following theorem.
%

\begin{theorem} \label{thm:cyclic_sets_from_Z}
Let $\rmatroid$ be a matroid with $0_\cflats = \emptyset$ and $1_\cflats = E$ where $\cflats = \cflats(M)$. Then, for any $X \subseteq E$, $X \in \cycles(M)$ if and only if the cyclic flat $$F^X=\bigwedge \{F\in\cflats : X\subseteq F\}$$ is such that $$\rho(F) + |X \setminus F| > \rho(F^X).
$$ for all $F \subsetneq F^X$ in $\cflats$.
\end{theorem}  

If this is indeed the case, then it is easy to verify that  $F^X$ as defined in Theorem \ref{thm:cyclic_sets_from_Z} is indeed the closure $\cl(X)$ as defined earlier. In order to analyze the parameters $\xnkd$ of a punctured code $C_X$, we will use the lattice of cyclic flats of $M_{C|X}$.
  
\begin{theorem} \label{thm:U_via_Z}
Let $\rmatroid$ be a matroid with $0_\cflats = \emptyset$ and $1_\cflats = E$ where $\cflats = \cflats(M)$. Then, for $X \in \cycles(M)$,
$$ 
\begin{array}{rl}
(i) & \cflats(M|X) = \{X \cap F \in \cycles(M): F \in \cflats \hcom F \subseteq F^X\},\\
(ii) & Y \in \cflats(M|X) \Rightarrow \rho_{|X}(Y) = \rho(F^Y).
\end{array}
$$
\end{theorem}

We remark that if $X$ is a cyclic flat of a matroid $M$, then
$\cflats(M|X) = \{F \in \cflats(M) : F \subseteq X\}$.

\begin{example} \label{ex:cyclic_sets_via_Z}
Let $\cflats = \cflats(M_C)$ be the lattice of cyclic flats given in Example \ref{ex:Z}, where $\cmatroid$ is the matroid associated to the linear LRC $C$ generated by the matrix $G$ given in Example \ref{ex:dss_code}. Then, $F^X = F^Y = Y_1$ for $X = \{1,2,3,7\}$ and $Y = \{1,2,3\}$. Further, $X$ is a cyclic set but $Y$ is not a cyclic set. The lattice of cyclic flats $(\cflats_X, \subseteq)$ for $M_{C|X} = M_C|X$ is shown in the following figure.

$$
\begin{tikzpicture}
    \node[shape=circle,draw=black] (00) at (0,0) {$0_{\cflats_X}$}; 
    \node[shape=circle,draw=black] (11) at (0,2) {$1_{\cflats_X}$};
    
    \node (t00) at (0,-0.8) { $(\emptyset,0)$ };  
    \node (t11) at (-0,2.8) { $(\{1,2,3,7\},3)$}; 
    
     \path [-] (00) edge (11);
\end{tikzpicture}
$$
\end{example}

The very simple structure of $\cflats_X$ shows that $C_X$ has the very favourable property of being an MDS code. Indeed, the following proposition is immediate from the definitions of the involved concepts.

\begin{proposition}
Let $C$ be a linear code of length $n$ and rank $k$. The following are equivalent:
\begin{enumerate}[(i)]
\item $C$ is an $[n,k,n-k+1]$-MDS code.
\item $M_C$ is the uniform matroid $U_n^k$.
\item $\cflats=\cflats(M_C)$ is the two element lattice with $1_\cflats=[n]$ and $0_\cflats=\emptyset$. 
\end{enumerate}
\end{proposition}

For linear LRCs we are also interested in when a coordinate set is an information set, or equivalently, if it is a basis for the matroid. This property is determined by the cyclic flats as follows.

\begin{proposition}~\label{infoset}
Let $C$ be a linear $\lcode$ with $0_\cflats = \emptyset$ and $1_\cflats = E$ where $\cflats$ is the collection of cyclic flats of the  matroid $\cmatroid$. Then, for any $X \subseteq E$, $X$ is an information set of $C$ if and only if the following two conditions are satisfied,
$$
\begin{array}{rl}
(i) & |X| = \rho_C(1_\cflats),\\
(ii) & |X \cap F| \leq \rho_C(F) \hbox{ for every } F \in \cflats.
\end{array}
$$
\end{proposition}

\begin{example}
Let $C$ be the linear $\nkd$-code generated by the matrix $G$ given in Example \ref{ex:dss_code}. Then, by the lattice of cyclic flats for $M_C$ given in Example \ref{ex:Z}, $C$ is a linear LRC with all-symbol $(2,2)$-locality. We notice that, by Proposition~\ref{infoset}, $\{1,2,3,4\}$ is an information set of $C$. This follows  as it is not contained in either of $Y_1$ and $Y_2$, while all its subsets are. On the other hand, $\{1,2,8,9\}$ is not an information set of $C$, as it is itself a subset of $Y_2$.
\end{example}

The parameters $\nkd$ of a linear LRC $C$ and $\xnkd$ of a punctured code $C_X$ that is a non-degenerate storage code can now be determined by the lattice of cyclic flats as follows.

\begin{theorem} \label{thm:nX_kX_dX_via_Z}
Let $C$ be a linear $\nkd$-LRC, where $\cflats = \cflats(M_C)$ for the matroid $\cmatroid$. Then, for any $X \in \cycles(M_C)$, $C_X$ is a linear $\xnkd$-LRC with
$$
\begin{array}{rlcl}
(i) & n_X &=& |X|,\\
(ii) & k_X &=& \rho(F^X),\\
(iii) & d_X &=& n_X - k_X + 1 -  \max \{\eta(Y) : Y \in \cflats(M_C|X) \setminus X\}.
\end{array}
$$ 
\end{theorem}

\begin{example}
Let $\cflats = \cflats(M_C)$ be the lattice of cyclic flats given in Example \ref{ex:Z}, where $\cmatroid$ is the matroid associated to the linear $\nkd$-LRC $C$ generated by the matrix $G$ given in Example \ref{ex:dss_code}. Then by Example \ref{ex:cyclic_sets_via_Z},  $X = \{1,2,3,7\}$ is a cyclic set, and by Theorem \ref{thm:nX_kX_dX_via_Z} $C_X$ is a linear $\xnkd$-LRC with parameters $n_X = 4$, $k_X = 3$ and $d_X = 4-3+1-0= 2$. Moreover, $n = n_E = 9$, $k = k_E = 4$ and $d = d_E = 9-4+1-3=3$.
\end{example}

\section{Singleton-type Bounds}
\label{sec:bounds}

Many important properties of a linear code $C$ are due to its matroid structure, which is captured by the matroid $M_C$. By the results in \cite{westerback15} and \cite{izs16}, matroid theory seems to be particularly suitable for proving  Singleton-type bounds for linear LRCs and nonexistence results of Singleton-optimal linear LRCs for certain parameters. 

Even though matroids can be associated to linear codes to capture the key properties for linear LRCs, this cannot be done in general  for nonlinear codes. Fortunately, by using some key properties of entropy, any code (either linear and nonlinear) $C$ can be associated with a polymatroid $P_C$ so that $P_C$ captures the key properties of the code when it is used as a LRC. A polymatroid is a generalisation of a matroid. For any linear code $C$ the associated polymatroid $P_C$ and matroid $M_C$ are the same object. We will briefly discuss the connection between polymatroids and codes in Section\ref{sec:beyond_linear_storage}.  Singleton-type bounds for polymatroids were derived in \cite{polymatroid}, and polymatroid theory for its part seems to be particularly  suitable for proving such bounds for general LRCs. In Section~\ref{sec:beyond_linear_storage} we will also review Singleton-type bounds for polymatroids and general codes with availability and hierarchy.

\subsection{Singleton-type Bounds for Matroids and Linear LRCs}

Matroid theory provides a unified way to understand and connect several different branches of mathematics, for example linear algebra, graph theory, combinatorics, geometry, topology and optimisation theory. Hence, a theorem proven for matroids gives ``for free'' theorems for many different objects related to matroid theory. As described earlier, the key parameters $(n,k,r,d,\delta,t)$ of a linear LRCs $C$ are matroid properties of the matroid $M_C$ and can therefore be defined for matroids in general. 

\begin{definition} \label{def:param_matroids}
Let $M = (\rho,E)$ be a matroid and $X \subseteq E$. Then
\begin{enumerate}[(i)]
\item $n_X = |X|$,
\item $k_X = \rho(X)$,
\item $d_X = \min \{|Y| : Y \subseteq X \hcom \rho(X \setminus Y) < \rho(X)\}$,
\item $X$ is an information set of $M$ if $\rho(X) = k_E$ and $\rho(Y) < k_E$ for all $Y \subsetneq X$,
\item $M$ is non-degenerate if $\rho(x) > 0$ for all $x \in E$ and $d_E \geq 2$.  
\end{enumerate}

Further, $n=n_E$, $k=k_E$ and $d=d_E$, and the definitions of repair sets, $(r,\delta)$-locality and $t$-availability for elements $x \in E$, as well as the concepts of $(n,k,d,r,\delta,t)_i$-matroids and $(n,k,d,r,\delta,t)_a$-matroids are directly carried over from Definitions \ref{def:LRC_local_avail} and \ref{def:s_i_a_LRC}. 
\end{definition}

 Before stating Theorem \ref{thm:bound_matroid_nkdrd} below, it is not at all clear that the Singleton-type bounds, already proven for linear LRCs, also hold for matroids in general. Especially, one could doubt this generality of the bound because of the wide connection between matroids and a variety of different mathematical objects, as well as for the sake of the recently proven result, stated in Theorem \ref{thm:almostall} later on, that almost all matroids are nonrepresentable. However, Theorem \ref{thm:bound_matroid_nkdrd} gives a Singleton-type bound that holds for matroids in general. This implies, as special cases, the same bound on linear LRCs and other objects related to matroids, {\em e.g.}, graphs, almost affine LRCs, and transversals. For the bound to make sense for various objects, a description of the parameters $(n,k,d,r,\delta)$ has to be given for the objects in question. To give an example, for a graph, 
 
\begin{itemize}
\item $n$ equals the number of edges,
\item $k$ equals the difference of the number of vertices and the number of connected components,
\item  $d$ is the smallest number of edges in an edge-cut (\emph{i.e.}, a set of edges whose removal increases the number of connected components in the graph).
\end{itemize}

Recall that the Singleton bound \cite{singleton64} states that for any linear $[n,k,d]$-code we have
\begin{equation} \label{eq:bound_linear_nkd}
d \leq n - k + 1.
\end{equation}
In what follows, we state generalised versions of this bound, accounting for the various parameters relevant for storage systems. We start with the general one for  matroids. 

\begin{theorem}[\cite{westerback15} Singleton-type bound for matroids] \label{thm:bound_matroid_nkdrd}
Let $M = (\rho,E)$ be an $(n,k,d,r,\delta)_i$-matroid. Then
\begin{equation} \label{eq:bound_matroid_nkdrd}
d \leq n - k + 1 - \left( \left \lceil \frac{k}{r} \right \rceil - 1 \right) (\delta - 1).
\end{equation}
\end{theorem}

Theorem \ref{thm:bound_matroid_nkdrd} was stated for all-symbol locality  in \cite{westerback15}. However, the proof given in \cite{westerback15} implies also information-symbol locality. As an illustration on how matroid theory and cyclic flats can be useful for proving Singleton-type of bounds we will here give a proof of Theorem \ref{thm:bound_matroid_nkdrd}. 
\begin{proof} We know from Theorem \ref{thm:nX_kX_dX_via_Z} that $d = n - k + 1 - \max \{\eta(Z) : Z \in \cflats \setminus E \}$. Hence to prove the theorem we need to show that there exists a cyclic flat $Z \neq E$ in $M$ with $\eta(Z) \geq \left( \left \lceil \frac{k}{r} \right \rceil - 1 \right) (\delta - 1)$. 

Let $B$ be an information set of $M$, \emph{i.e,} $B$ is a basis of $M$, such that $M$ is an $(n,k,d,r,\delta)_i$-matroid. For $x \in B$ let $R_x$ denote the repair set of $x$. Since $R_x$ is a cyclic set of  we obtain that $Z_x = \cl(R_x)$ is a cyclic flat of $M$ with 
$$
\rho(Z_x) = \rho(R_x) \leq r \hbox{ and } \eta(Z_x) \geq \eta(R_x) \geq \mathrm{d}_{R_x} - 1 \geq \delta - 1.
$$

As $\rho(B) = k$ we can choose a subset of cyclic flats $\{Z_1, \ldots, Z_m\} \subseteq \{Z_x : x \in B\}$ such that we obtain a chain of cyclic flats
$$
\emptyset = Y_0 \subsetneq Y_1 \subsetneq \ldots \subsetneq Y_m = E
$$
with $Y_i = Y_{i-1} \vee Z_i$ for $i = 1,\ldots,m$. Since $\rho(Y_0) = \eta(Y_0) = 0$ and $\rho(Y_m) = k$, the theorem will now be proved if we can prove that $\rho(Y_i)-\rho(Y_{i-1}) \leq r$ and $\eta(Y_i)-\eta(Y_{i-1}) \geq \delta - 1$ for $i = 1,\ldots,m$. 

First, by the use of Axiom (R.3) and Proposition \ref{prop:basic-Z}, 
$$
\begin{array}{rcl}
\rho(Y_i) & = & \rho(\cl(Y_{i-1} \cup Z_i)) = \rho(Y_{i-1} \cup Z_i) \leq \rho(Y_{i-1}) + \rho(Z_i) - \rho(Y_{i-1} \cap Z_i) \\
& \leq & \rho(Y_{i-1}) + r.
\end{array}
$$

Second, by Axiom (R.3), $\eta(X) + \eta(Y) \leq \eta(X \cap Y) + \eta(X \cup Y)$ for $X,Y \subseteq E$. Further, we observe that, $\mathrm{cyc}(Y_{i-1} \cap Z_i)$ and $Z_i$ are cyclic flats of $M|Z_i$ and that $\mathrm{cyc}(Y_{i-1} \cap Z_i) \subsetneq Z_i$. Hence,
$$
\begin{array}{rcl}
\eta(Y_i)&=& \eta(\cl(Y_{i-1} \cup Z_i)) \geq \eta(Y_{i-1} \cup Z_i) \geq \eta(Y_{i-1}) + \eta(Z_i) - \eta(Y_{i-1} \cap Z_i) \\
 &=& \eta(Y_{i-1}) + \eta(Z_i) - \eta(\mathrm{cyc}(Y_{i-1} \cap Z_i)) \\
 &=& \eta(Y_{i-1}) + |Z_i| - \rho(Z_i) -  \eta(\mathrm{cyc}(Y_{i-1} \cap Z_i)) \geq \eta(Y_{i-1}) + d_{Z_i} - 1\\
 &\geq& \eta(Y_{i-1}) + \delta - 1.
\end{array}
$$
That $d_{Z_i} \geq \delta$ follows from the fact that $Z_i = \cl(R_x)$ for some $x \in  B$ and therefore
$$
\begin{array}{rcl}
d_{Z_i} = d_{\cl(R_x)} &=& \min \{|Y| : Y \subseteq \cl(R_x) \hbox{, } \rho(\cl(R_x) \setminus Y) \leq \rho(\cl(R_x)\} \\
&\geq& \min \{|Y| : Y \subseteq R_x \hbox{, } \rho(R_x \setminus Y) \leq \rho(R_x)\} \\
&=& d_{R_x}.
\end{array}
$$
$\hfill\qed$
\end{proof}

The Singleton bound given in (\ref{eq:bound_linear_nkd}) was generalised by Gopalan \emph{et al.} in \cite{gopalan12} as follows. A linear $(n,k,d,r)_i$-LRC satisfies 
\begin{equation} \label{eq:bound_linear_nkdr}
d \leq n - k + 1 - \left(  \left \lceil\frac{k}{r} \right \rceil - 1 \right).
\end{equation}
The bound (\ref{eq:bound_linear_nkdr}) shows that there is a penalty for requiring locality. That is, the smaller the locality $r$ the smaller the upper bound on $d$. By the definition of LRCs, any linear $[n,k,d]$-code with locality $r$ is also a linear $[n,k,d]$-code with locality $k$. Hence, by letting the locality be $k$, the bound (\ref{eq:bound_linear_nkdr}) implies  (\ref{eq:bound_linear_nkd}).

The bound (\ref{eq:bound_linear_nkdr}) was generalised in \cite{prakash12} as follows. A linear $(n,k,d,r,\delta)_i$-LRC satisfies
\begin{equation} \label{eq:bound_linear_nkdrd}
d \leq n - k + 1 - \left(  \left \lceil\frac{k}{r} \right \rceil - 1 \right)(\delta - 1).
\end{equation}
The bound $(\ref{eq:bound_linear_nkdrd})$ again shows that there is a penalty on the upper bound for $d$ depending on the size of the local distance $\delta$. This is, the bigger the local distance $\delta$ the smaller the upper bound on the global distance $d$. However, we must remark that any linear $(n,k,d,r,\delta)_i$-LRC satisfies $d \geq \delta$, and this property also holds more generally for matroids \cite{westerback15}. The bound (\ref{eq:bound_linear_nkdr}) follows from the bound (\ref{eq:bound_linear_nkdrd}) by letting $\delta=2$.

A bound including availability was proven in \cite{wang14b}. This bound states that a linear $(n,k,d,r,t)_i$-LRC satisfies
\begin{equation} \label{eq:bound_linear_nkdrt}
d \leq n - k + 1 - \left(  \left \lceil\frac{t(k-1)+1}{t(r-1)+1} \right \rceil - 1 \right).
\end{equation}
Again, the bound (\ref{eq:bound_linear_nkdr}) follows from  (\ref{eq:bound_linear_nkdrt}) above by letting $t=1$. 

The bounds  (\ref{eq:bound_matroid_nkdrd})-(\ref{eq:bound_linear_nkdrt}) are stated assuming information-symbol locality. However, since every matroid contains an information set, this implies that the bound is also  valid under the stronger assumption of all-symbol locality.

\subsection{Stronger Bounds for Certain Parameter Values}

A linear LRC, or more generally a matroid, that achieves any of the Singleton-type bounds given above will henceforth be called \emph{Singleton-optimal}. 

Any $(n,k,d,r,\delta)_i$-matroid $M$ satisfies that $\delta \leq d$. Hence, by the bound (\ref{eq:bound_linear_nkdrd}), $k \leq n - (\delta-1)\lceil \frac{k}{r} \rceil$ for $M$. Thus, regardless of the global minimum distance $d$, any $(n,k,d,r,\delta)-$LRC with either information or all-symbol locality, has parameters $n,k,r,\delta$ in the set
\begin{equation} \label{eq:possible_parameters}
P(n,k,r,\delta) = \left\{(n,k,r,\delta) \in \mathbb{Z}^4 : 2 \leq \delta \hbox{ and } 0 < r \leq k \leq  n - (\delta-1) \left \lceil \frac{k}{r} \right \rceil \right\}. 
\end{equation}
A very natural question to ask then is for which parameters $(n,k,r,\delta) \in P(n,k,r,\delta)$  there exists a Singleton-optimal matroid or linear LRC, regarding both information and all-symbol locality. We remark that existence results on Singleton-optimal linear LRCs imply existence results on Singleton-optimal matroids. Conversely, nonexistence results on Singleton-optimal matroids implies nonexistence results on Singleton-optimal linear LRCs.

When considering information-symbol locality it is known that the upper bound for $d$ given in (\ref{eq:bound_linear_nkdrd}) is achieved for all parameters $(n,k,r,\delta) \in P(n,k,r,\delta)$ by linear LRCs over sufficient large fields. This follows from \cite{huang07}, where a new class of codes called pyramid codes was given. Using this class of codes, Singleton-optimal linear $(n,k,d,r,\delta)_i$-LRCs can be constructed for all parameters in $P(n,k,r,\delta)$.

It is well known that Singleton-optimal linear $(n,k,d,r,\delta)_a$-LRCs exist when $r=k$. Namely, the LRCs in these cases are  linear $[n,k,n-k+1]$ MDS-codes. However, existence or nonexistence results when $r<k$ are in general not that easy to obtain. In \cite{song14}, existence and nonexistence results on Singleton-optimal linear $(n,k,d,r,\delta)_a$-LRCs were examined. Such results were given for certain regions of parameters, leaving other regions for which the answer of existence or nonexistence of Singleton-optimal linear LRCs is not known. The results on nonexistence were extended to matroids in $\cite{westerback15}$. All the  parameter regions for the nonexistence of Singleton-optimal linear LRCs in \cite{westerback15} were also regions of parameters for the nonexistence of Singleton optimal matroids for all-symbol locality. Further, more regions of parameters for nonexistence of Singleton-optimal matroids with all-symbol locality were given in \cite{westerback15}. This implies new regions of parameters for nonexistence of Singleton-optimal linear LRCs with all-symbol locality. 


The nonexistence results for Singleton-optimal matroids were proven via the following structure result in \cite{westerback15}. 
Before we state the theorem we need the concept of nontrivial unions. Let $M = (\rho,E)$ be a matroid with repair sets $\{R_x\}_{x \in E}$. For $Y \subseteq E$, we say that
$$
R_Y = \bigcup_{x \in Y} R_x \quad \hbox{and} \quad R_Y \hbox{ is a \emph{nontrivial union} if $R_x \nsubseteq R_{Y \setminus \{x\}}$ for every $x \in Y$.} 
$$

\begin{theorem} [\cite{westerback15} Structure theorem for Singleton-optimal matroids] \label{thm:matroid_structure}
Let $M = (\rho, E)$ be an $(n,k,d,r,\delta)_a$-matroid with $r < k$, repair sets $\{R_x\}_{x \in E}$ and 
$$
d = n - k + 1 - \left(  \left \lceil\frac{k}{r} \right \rceil - 1 \right)(\delta - 1).
$$
Then, the following properties must be satisfied by the collection of repair sets and the lattice of cyclic flats $\mathcal{Z}$ of $M$:
$$
\begin{array}{cl}
(i) & 0_{\mathcal{Z}} = \emptyset,\\
(ii) & \hbox{for each } x \in E,\\
 & 
 \begin{array}{ll}
 a) & R_x \hbox{ is an atom of } \mathcal{Z} \hbox{, }\\
    &  (i.e.\hbox{, } R_x \in \cflats \hbox{, } 0_\cflats < R_x \hbox{ and } \nexists Z \in \cflats \hbox{ such that } 0_\cflats < Z < R_x),\\
 b) & \eta(R_x) = \delta - 1,
 \end{array}\\
 (iii) & \hbox{for each $Y \subseteq E$ with $R_Y$ being a nontrivial union,} \\
 & 
 \begin{array}{ll}
 c) & |Y| < \lceil \frac{k}{r} \rceil \quad \Rightarrow \quad R_Y \in \mathcal{Z},\\
  d) & |Y| < \lceil \frac{k}{r} \rceil \quad \Rightarrow \quad \rho(R_Y) = |R_Y| - |Y|(\delta - 1),\\
  e) & |Y| \leq \lceil \frac{k}{r} \rceil \quad \Rightarrow \quad 
  |R_x \bigcap (R_{Y \setminus \{x\}}) | \leq |R_x| - \delta, \hbox{ for each } x \in Y,\\
  f) & |Y| \geq \lceil \frac{k}{r} \rceil \quad \Rightarrow \quad \{Z \in \mathcal{Z} : Z \supseteq R_Y \} = 1_{\mathcal{Z}} = E.
 \end{array}
\end{array}
$$ 
\end{theorem}

Conditions (i) and (ii) in the structure theorem above for Singleton-optimal matroids show that each repair set $R_x$ must correspond to a uniform matroid with $|R_x|$ elements and rank $|R_x| - (\delta-1)$. Further, condition (iii) gives structural properties on nontrivial unions of repair sets. This can be viewed as structural conditions on how nontrivial unions of uniform matroids need to be glued together in a Singleton-optimal matroid, with the uniform matroids corresponding to repair sets. For Singleton-optimal linear $(n,k,d,r,\delta)_a$-LRCs, the property of the repair sets being uniform matroids corresponds to the repair sets being linear $[|R_x|, |R_x|- (\delta - 1), \delta]$-MDS codes. We remark that structure theorems when $r|k$ for Singleton-optimal linear $(n,k,d,r,\delta)_a$-LRCs and the special case of $(n,k,d,r,\delta=2)_a$-LRCs are given in \cite{kamath14} and \cite{gopalan12}, respectively. These theorems show that the local repair sets correspond to linear  $[r+\delta-1, r, \delta]$-MDS codes that are mutually disjoint. This result is a special case of Theorem \ref{thm:matroid_structure}.  

\begin{example} \label{ex:singleton_optimal_matroid}
By Theorem \ref{th:Z-axiom}, the poset with its associated subsets of $E = [16]$ and rank of these subsets in the figure below defines the set of cyclic flats $\cflats$ and the rank function restricted to the sets in $\cflats$ of a matroid $M$ on $E$. From Theorem \ref{thm:nX_kX_dX_via_Z}, we obtain that $(n,k,d) = (16,7,6)$ for the matroid $M$. Choosing repair sets $R_1 = \cdots = R_5 = X_1$, $R_6 = \cdots = R_9 = X_2$, $R_{10} = \cdots = R_{13} = X_3$ and $R_{14} = \cdots = R_{16} = X_4$,
we obtain that $M$ is an $(n=16,k=7,d=6,r=3,\delta=3)_a$-matroid. It can easily be checked that all the properties (i)-(iii) are satisfied by the matroid $M$ and the chosen repair sets. Further, we also have that $M$ is Singleton-optimal as
$$
n - k + 1 - \left(  \left \lceil\frac{k}{r} \right \rceil - 1 \right)(\delta - 1) = 6 = d.
$$
$$
\begin{tikzpicture}
    \node[shape=circle,draw=black] (00) at (0,0) {$0_\cflats$};  
    \node[shape=circle,draw=black] (11) at (-2.7,2) {$X_1$}; 
    \node[shape=circle,draw=black] (12) at (-0.9,2) {$X_2$}; 
    \node[shape=circle,draw=black] (13) at (0.9,2) {$X_3$}; 
    \node[shape=circle,draw=black] (14) at (2.7,2) {$X_4$};  
    \node[shape=circle,draw=black] (21) at (-4.5,4) {$Y_1$}; 
    \node[shape=circle,draw=black] (22) at (-2.7,4) {$Y_2$}; 
    \node[shape=circle,draw=black] (23) at (-0.9,4) {$Y_3$}; 
    \node[shape=circle,draw=black] (24) at (0.9,4) {$Y_4$}; 
    \node[shape=circle,draw=black] (25) at (2.7,4) {$Y_5$}; 
    \node[shape=circle,draw=black] (26) at (4.5,4) {$Y_6$}; 
    \node[shape=circle,draw=black] (31) at (0,6) {$1_\cflats$};
    
      \path [ultra thin] (00) edge (11);
     \path [ultra thin]  (00) edge (12);
     \path [ultra thin]  (00) edge (13);
     \path [ultra thin]  (00) edge (14);
     \path [ultra thin]  (11) edge (21);
     \path [ultra thin]  (11) edge (22);
     \path [ultra thin]  (11) edge (23);
     \path [ultra thin]  (12) edge (21);
     \path [ultra thin]  (12) edge (24);
     \path [ultra thin]  (12) edge (25);
     \path [ultra thin]  (13) edge (22);
     \path [ultra thin]  (13) edge (24);
     \path [ultra thin]  (13) edge (26);
     \path [ultra thin]  (14) edge (23);
     \path [ultra thin]  (14) edge (25);
     \path [ultra thin]  (14) edge (26);
     \path [ultra thin]  (21) edge (31);
     \path [ultra thin]  (22) edge (31);
     \path [ultra thin]  (23) edge (31);
     \path [ultra thin]  (24) edge (31);
     \path [ultra thin]  (25) edge (31);
     \path [ultra thin]  (26) edge (31);
    
    \node [fill=white,rounded corners=2pt,inner sep=1pt] (t00) at (0,-0.7) { \small{$ (\emptyset,0) $ }}; 
    \node [fill=white,rounded corners=2pt,inner sep=1pt] (t11) at (-2.7,1.4) {\small{ $ ( \{1-5\},3) $ }}; 
    \node [fill=white,rounded corners=2pt,inner sep=1pt] (t12) at (-0.9,2.6) {\small{ $ (\{5-9\},3) $ }};
    \node [fill=white,rounded corners=2pt,inner sep=1pt] (t13) at (0.8,1.4) {\small{ $( \{9-13\},3) $ }};
    \node [fill=white,rounded corners=2pt,inner sep=1pt] (t14) at (2.7,2.6) {\small{ $ ( \{1, 13-16\},3) $ }};     
    \node [fill=white,rounded corners=2pt,inner sep=1pt] (t21) at (-4.5,4.6) {\small{ $ ( \{1-9\},5)$ }};
    \node [fill=white,rounded corners=2pt,inner sep=1pt] (t22) at (-2.7,3.4) {\small{ $ ( \{1-5, 9-13\},6)$ }};
    \node [fill=white,rounded corners=2pt,inner sep=1pt] (t23) at (-0.9,4.6) {\small{ $( \{1-5, 13-16\},5) $ }};
    \node [fill=white,rounded corners=2pt,inner sep=1pt] (t24) at (0.9,3.4) {\small{ $ ( \{5-13\},5) $ }};
    \node [fill=white,rounded corners=2pt,inner sep=1pt] (t25) at (2.7,4.6) {\small{ $ ( \{1,5-9, 13-16\},6) $ }};
    \node [fill=white,rounded corners=2pt,inner sep=1pt] (t26) at (4.5,3.4) {\small{ $ ( \{1, 9-16\},5) $ }};
    \node [fill=white,rounded corners=2pt,inner sep=1pt] (t31) at (0,6.7) {\small{ $ ( [16],7) $ }};

\end{tikzpicture}
$$
\end{example}

\section{Code Constructions}
\label{sec:constructions}

\subsection{Constructions of $(n,k,d,r,\delta)_a$-matroids via Cyclic Flats}
In \cite{westerback15}, a construction of a broad class of linear $(n,k,d,r,\delta)_a$-LRCs is given via matroid theory. This is generalised in \cite{polymatroid} and \cite{izs16} to account for availability and hierarchy, respectively. 

\emph{A construction of $(n,k,d,r,\delta)_a$-matroids via cyclic flats:}\\ 
Let $F_1,\ldots,F_m$ be a collection of finite sets and $E = \bigcup_{i=1}^m F_i $. Assign a function $\rho:\{F_i\}\cup\{E\} \rightarrow \mathbb{Z}$ satisfying
\begin{equation} \label{eq:matroid_construction_conditions}
\begin{array}{cl}
(i) & 0 < \rho(F_i) < |F_i| \hbox{ for } i \in [m],\\
(ii) & \rho(F_i) < \rho(E) \hbox{ for } i \in [m],\\
(iii) & \rho(E) \leq |E| - \sum_{i \in [m]}\eta(F_i),\\
(iv) & j \in [m] \Rightarrow |F_{[m] \setminus \{j\}} \cap F_j| < \rho(F_j),
\end{array}
\end{equation}
where
$$
\eta(F_i) = |F_i| - \rho(F_i) \hbox{ for } i \in [m] \quad \hbox{and} \quad F_I = \bigcup_{i \in I} F_i \hbox{ for } I \subseteq [m].
$$
Extend $\rho$ to $\{F_I\}\to\mathbb{Z}$ by
\begin{equation} \label{eq:matroid_construction_rank}
\rho(F_I)  = \min \{ |F_I| - \sum_{i \in I} \eta(F_i),\rho(E) \}
\end{equation}
and let $\mathcal{Z}$ be the following collection of subsets of $E$,
\begin{equation} \label{eq:matroid_construction_Z}
\mathcal{Z} = \{F_I : I \subseteq [m] \hbox{ and } \rho(F_I) < \rho(E)\} \cup E.
\end{equation}

\begin{theorem} [\cite{westerback15} Construction of $(n,k,d,r,\delta)_a$-matroids] \label{thm:matroid_construction}
Let $F_1,\ldots,F_m,$ be a collection of finite sets with $E = \bigcup_{i=1}^m F_i $ and $\rho:\{F_i\}_{i \in [m]} \rightarrow \mathbb{Z}$ satisfying  (\ref{eq:matroid_construction_conditions}). Then the pair $(\rho, \mathcal{Z})$, defined in (\ref{eq:matroid_construction_rank}) and (\ref{eq:matroid_construction_Z}), defines a $(n,k,d,r,\delta)_a$-matroid $M(F_1,\ldots,F_m; \rho)$ on $E$ for which $\mathcal{Z}$ is the collection of cyclic flats, $\rho$ is the rank function restricted to the cyclic flats in $\mathcal{Z}$, $F_1,\ldots,F_m,$ are the repair sets and
$$
\begin{array}{cl}
(i) & n = |E|,\\
(ii) & k = \rho(E),\\
(iii) & d = n - k +1 - \max \{\sum_{i \in I } \eta(F_i) : F_I \in \mathcal{Z} \setminus E \},\\
(iv) & \delta = 1 + \min \{\eta(F_i) : i \in [m]\},\\
(v) & r = \max \{\rho(F_i) : i \in {m} \}.
\end{array}
$$
\end{theorem}

That $M(F_1,\ldots,F_m;\rho)$ defines a matroid follows from a proof given in \cite{westerback15} that the pair $(\rho,\mathcal{Z})$ satisfies the axiomatic scheme of matroids via cyclic flats and their ranks stated in Theorem \ref{th:Z-axiom}. The correctness of the parameters $(n,k,d,r,\delta)$ when $F_1,\ldots,F_m$ are considered as the repair sets also follows from \cite{westerback15}.

We remark, that the matroids constructed in Theorem \ref{thm:matroid_construction} satisfy, for all unions of repair sets $F_I$ with $\rho(F_I) < \rho(E)$, that
\begin{equation} \label{eq:property_optimal}
\begin{array}{cl}
(i) & \hbox{$F_I$ is a cyclic flat},\\
(ii)& \hbox{the nullity $\eta(F_I)$ of $F_I$  is as small as possible}.\\
\end{array}
\end{equation}
 Properties (i) and (ii) above are trivially seen to be fulfilled by uniform matroids $U_{k,n}$, where $\mathcal{Z} = \{\emptyset, E\}$, $\rho(\emptyset) = 0$  and $\rho(E) = k$. However, uniform matroids cannot be constructed by Theorem \ref{thm:matroid_construction}, since all constructed matroids by this theorem have $r<k$ and uniform matroids have $r=k$. Though both uniform matroids and the matroids constructed in Theorem \ref{thm:matroid_construction} satisfy properties (i) and (ii) in (\ref{eq:property_optimal}), we will consider them in terms of a class of matroids $\mathcal{M}$, defined as follows:
\begin{equation} \label{eq:matroid_class}
\mathcal{M} = \{M = M(F_1,\ldots,F_m;\rho) : M \hbox{ is constructed in Theorem \ref{thm:matroid_construction}}\} \cup \{U_{k,n}\}.
\end{equation}
By the structure Theorem \ref{thm:matroid_structure}, the properties (i) and (ii) in (\ref{eq:property_optimal}) are necessary (but not sufficient) for Singleton-optimal $(n,k,d,r,\delta)_a$-matroids.

\begin{example} \label{ex:construction_Z}
 Let $E = [12]$ and let $F_1 = \{1,\ldots,4\}$, $F_2 = \{3,\ldots,6\}$, $F_3 = \{7,\ldots,10\}$, $F_4 = \{10,\ldots,12\}$ with $\rho(F_1) = \rho(F_2) = \rho(F_3) = 3$, $\rho(F_4) = 2$, and $\rho(E) = 7$. Then, by Theorem \ref{thm:matroid_construction}, $M(F_1,\ldots,F_m;\rho)$ is an $(n,k,d,r,\delta)_a$-matroid over $E$ with $(n,k,d,r,\delta) = (12,7,3,3,2)$ and the following lattice of cyclic flats and their ranks. 

$$
\begin{tikzpicture}
    \node[shape=circle,draw=black] (00) at (0,0) {\small$\rho(\emptyset)=0$};  
    \node[shape=circle,draw=black] (11) at (-2.7,2) {\small$\rho(F_1)=3$}; 
    \node[shape=circle,draw=black] (12) at (-0.9,2) {\small$\rho(F_2)=3$}; 
    \node[shape=circle,draw=black] (13) at (0.9,2) {\small$\rho(F_3)=3$}; 
    \node[shape=circle,draw=black] (14) at (2.7,2) {\small$\rho(F_4)=2$};  
    \node[shape=circle,draw=black] (21) at (-4.5,4) {\small$\rho=4$}; 
    \node[shape=circle,draw=black] (22) at (-2.7,4) {\small$\rho=6$}; 
    \node[shape=circle,draw=black] (23) at (-0.9,4) {\small$\rho=5$}; 
    \node[shape=circle,draw=black] (24) at (0.9,4) {\small$\rho=6$}; 
    \node[shape=circle,draw=black] (25) at (2.7,4) {\small$\rho=5$}; 
    \node[shape=circle,draw=black] (26) at (4.5,4) {\small$\rho=4$}; 
    \node[shape=circle,draw=black] (31) at (-2.7,6) {\small$\rho=6$}; 
    \node[shape=circle,draw=black] (41) at (0,7) {\small$\rho(1_\cflats)=7$};
    
    
     \path [-] (00) edge (11);
     \path [-] (00) edge (12);
     \path [-] (00) edge (13);
     \path [-] (00) edge (14);
     \path [-] (11) edge (21);
     \path [-] (11) edge (22);
     \path [-] (11) edge (23);
     \path [-] (12) edge (21);
     \path [-] (12) edge (24);
     \path [-] (12) edge (25);
     \path [-] (13) edge (22);
     \path [-] (13) edge (24);
     \path [-] (13) edge (26);
     \path [-] (14) edge (23);
     \path [-] (14) edge (25);
     \path [-] (14) edge (26);
     \path [-] (21) edge (31);
     \path [-] (23) edge (31);
     \path [-] (25) edge (31);
     \path [-] (31) edge (41);
     \path [-] (22) edge (41);
     \path [-] (24) edge (41);
     \path [-] (26) edge (41);
\end{tikzpicture}
$$
Further, the matroid is not Singleton-optimal since
$$
d = 3 < n-k+1-\left ( \left \lceil \frac{k}{r} \right \rceil - 1 \right )(\delta - 1) = 4.
$$
\end{example}

\subsection{A Matroidal Construction of Linear All Symbol LRCs}

As will be explained below, all matroids constructed in Theorem \ref{thm:matroid_structure} are contained in a class of matroids called gammoids. These matroids are linear, which especially implies that all $(n,k,d,r,\delta)_a$-matroids constructed by Theorem \ref{thm:matroid_structure} are matroids associated with linear $(n,k,d,r,\delta)_a$-LRCs. 

\begin{definition}
Any (finite) directed graph $\Gamma = (V,D)$ and vertex subsets $E,T \subseteq V$ define a \emph{gammoid} $\gammoid$, where $\igammoid$ is a the matroid with 
$$
\indeps = \{X \subseteq E : \exists \hbox{ a set of $|X|$ vertex-disjoint paths from $X$ to $T$}\}.
$$ 	
\end{definition}

\begin{theorem}[\cite{lindstrom73}] \label{thm:gammoid_is_linear}
Every gammoid $\gammoid$ is $\field{q}$-linear for all prime powers 	$q \geq 2^{|E|}$.
\end{theorem}

In~\cite{westerback15}, it is proven that the matroids constructed in Theorem~\ref{thm:matroid_construction} are indeed gammoids, and hence representable. This is achieved by explicitly constructing a triple $(\Gamma, E, T)$ whose associated matroid is $M(F_1,\ldots,F_m;\rho)$. The details of the construction are left to Theorem~\ref{thm:setmatroid_equal_gammoid} in the appendix. The essence of the argument is to construct a graph of depth three, whose sources correspond to the ground set of the matroid, and whose middle layer corresponds to the repair sets, with multiplicities to reflect the ranks of the repair sets.

\begin{example} \label{ex:Z_to_gammoid}
The following directed graph $\Gamma = (V = E \cup H \cup T, D)$ is constructed in Theorem \ref{thm:setmatroid_equal_gammoid} from the matroid $\setmatroid$ given in Example \ref{ex:construction_Z}.
$$
\begin{tikzpicture}

\node (E) at (-4,0) { E: };
\node (1) at (-3,0) { 1 };
\node (2) at (-2.4,0) { 2 };
\node (3) at (-1.6,0) { 3 };
\node (4) at (-0.8,0) { 4 };
\node (5) at (0,0) { 5 };
\node (6) at (0.8,0) { 6 };
\node (7) at (1.6,0) { 7 };
\node (8) at (2.4,0) { 8 };
\node (9) at (3.2,0) { 9 };
\node (10) at (4,0) { 10 };
\node (11) at (4.8,0) { 11 };
\node (12) at (5.6,0) { 12 };

\node (H) at (-4,2) { H: };
\node (h1) at (-3,2) { $R_1$ };
\node (h2) at (-1.7,2) { $R_{\{1,2\}}$ };
\node (h3) at (0,2) { $R_{\{1,2\}}$ };
\node (h4) at (1,2) { $R_2$ };
\node (h5) at (2,2) { $R_3$ };
\node (h6) at (3,2) { $R_3$ };
\node (h7) at (4.3,2) { $R_{\{3,4\}}$ };
\node (h8) at (5.6,2) { $R_4$ };

\node (T) at (-4,4) { T: };
\node (t1)[shape=circle,fill=black] at (-3,4) {};
\node (t2)[shape=circle,fill=black] at (-1.5,4) {};
\node (t3)[shape=circle,fill=black] at (-0.1,4) {};
\node (t4)[shape=circle,fill=black] at (1.3,4) {};
\node (t5)[shape=circle,fill=black] at (2.7,4) {};
\node (t6)[shape=circle,fill=black] at (4.1,4) {};
\node (t7)[shape=circle,fill=black] at (5.6,4) {};

\path [->] (1) edge (h1);
\path [->] (1) edge (h2);
\path [->] (1) edge (h3);
\path [->] (2) edge (h1);
\path [->] (2) edge (h2);
\path [->] (2) edge (h3);
\path [->] (3) edge (h2);
\path [->] (3) edge (h3);
\path [->] (4) edge (h2);
\path [->] (4) edge (h3);
\path [->] (5) edge (h2);
\path [->] (5) edge (h3);
\path [->] (5) edge (h4);
\path [->] (6) edge (h2);
\path [->] (6) edge (h3);
\path [->] (6) edge (h4);
\path [->] (7) edge (h5);
\path [->] (7) edge (h6);
\path [->] (7) edge (h7);
\path [->] (8) edge (h5);
\path [->] (8) edge (h6);
\path [->] (8) edge (h7);
\path [->] (9) edge (h5);
\path [->] (9) edge (h6);
\path [->] (9) edge (h7);
\path [->] (10) edge (h7);
\path [->] (11) edge (h7);
\path [->] (11) edge (h8);
\path [->] (12) edge (h7);
\path [->] (12) edge (h8);
\path [->] (h1) edge (t1);
\path [->] (h1) edge (t2);
\path [->] (h1) edge (t3);
\path [->] (h1) edge (t4);
\path [->] (h1) edge (t5);
\path [->] (h1) edge (t6);
\path [->] (h1) edge (t7);
\path [->] (h2) edge (t1);
\path [->] (h2) edge (t2);
\path [->] (h2) edge (t3);
\path [->] (h2) edge (t4);
\path [->] (h2) edge (t5);
\path [->] (h2) edge (t6);
\path [->] (h2) edge (t7);
\path [->] (h3) edge (t1);
\path [->] (h3) edge (t2);
\path [->] (h3) edge (t3);
\path [->] (h3) edge (t4);
\path [->] (h3) edge (t5);
\path [->] (h3) edge (t6);
\path [->] (h3) edge (t7);
\path [->] (h4) edge (t1);
\path [->] (h4) edge (t2);
\path [->] (h4) edge (t3);
\path [->] (h4) edge (t4);
\path [->] (h4) edge (t5);
\path [->] (h4) edge (t6);
\path [->] (h4) edge (t7);
\path [->] (h5) edge (t1);
\path [->] (h5) edge (t2);
\path [->] (h5) edge (t3);
\path [->] (h5) edge (t4);
\path [->] (h5) edge (t5);
\path [->] (h5) edge (t6);
\path [->] (h5) edge (t7);
\path [->] (h6) edge (t1);
\path [->] (h6) edge (t2);
\path [->] (h6) edge (t3);
\path [->] (h6) edge (t4);
\path [->] (h6) edge (t5);
\path [->] (h6) edge (t6);
\path [->] (h6) edge (t7);
\path [->] (h7) edge (t1);
\path [->] (h7) edge (t2);
\path [->] (h7) edge (t3);
\path [->] (h7) edge (t4);
\path [->] (h7) edge (t5);
\path [->] (h7) edge (t6);
\path [->] (h7) edge (t7);
\path [->] (h8) edge (t1);
\path [->] (h8) edge (t2);
\path [->] (h8) edge (t3);
\path [->] (h8) edge (t4);
\path [->] (h8) edge (t5);
\path [->] (h8) edge (t6);
\path [->] (h8) edge (t7);

\end{tikzpicture}
$$
 \end{example}

In general it is extremely hard to prove that a matroid is linear (or the converse). There is no known deterministic algorithm to solve this problem in general. However, by combining the results given in Theorems \ref{thm:matroid_construction}--\ref{thm:setmatroid_equal_gammoid}, we obtain the following result.

\begin{theorem}[\cite{westerback15} \label{thm:matroid_to_lrc}
A matroidal construction of $(n,k,d,r,\delta)_a$-LRCs] \label{thm:setmatroid_equal_M_C}
For every $(n,k,d,r,\delta)_a$-matroid $\setmatroid$ given by Theorem \ref{thm:matroid_construction} and every prime power $q \geq 2^{|E|}$ there is a linear $(n,k,d,r,\delta)_a$-LRC $C$ over $\field{q}$ with repair sets $F_1,\ldots, F_m$ such that $\setmatroid = M_C$.
\end{theorem}

\begin{example}\label{12-7-code}
The $(12,7,3,3,2)_a$-matroid $\setmatroid$ given in Example \ref{ex:construction_Z} equals the matroid $M_C = M[G]$, where $G$ equals the following matrix over $\field{5}$:

$$
\begin{small}
G\,= \,
\begin{tabular}{ |c|c|c|c|c|c|c|c|c|c|c|c| }
\multicolumn{1}{c}{\,1\,}&\multicolumn{1}{c}{\,2\,}&\multicolumn{1}{c}{\,3\,}&\multicolumn{1}{c}{\,4\,}&\multicolumn{1}{c}{\,5\, }&\multicolumn{1}{c}{\,6\,}&\multicolumn{1}{c}{ \,7\,}&\multicolumn{1}{c}{\,8\, }&\multicolumn{1}{c}{\,9\, }&\multicolumn{1}{c}{10}&\multicolumn{1}{c}{11}&\multicolumn{1}{c}{12}\\
\hline
1&1&0&0&0&0&0&0&0&0&1&3\\
\hline
0&2&1&0&0&1&0&0&0&0&1&3\\
\hline
0&3&0&1&0&3&0&0&0&0&1&3\\
\hline
0&0&0&0&1&2&0&0&0&0&1&3\\
\hline
0&0&0&0&0&0&1&0&1&0&1&3\\
\hline
0&0&0&0&0&0&0&1&2&0&1&3\\
\hline
0&0&0&0&0&0&0&0&3&1&1&1\\
\hline
\end{tabular}
\end{small}
$$
Hence, the code $C$ generated by the rows of $G$ is a linear $(12,7,3,3,2)$-LRC over $\field{5}$ with repair sets $F_1 = \{1,2,3,4\}$, $F_2 = \{3,4,5,6\}$, $F_1 = \{7,8,9,10\}$ and $F_1 = \{10,11,12\}$. 

\end{example}

Note that the bound $q \geq 2^{|E|}$ given in Theorem \ref{thm:setmatroid_equal_M_C} is a very rough bound. There are many matroids $\setmatroid = M_C$ for linear LRCs $C$ over $\field{q}$ where $q \ll 2^{|E|}$. In Example~\ref{12-7-code}, for instance, we constructed a code over $\F_5$, while the field size predicted by Theorem \ref{thm:setmatroid_equal_M_C} was $2^{12}=4096 \gg 5$. To construct an explicit linear $(n,k,d,r,\delta)_a$-LRC from a matroid $\setmatroid$, one can use the directed graph representation of the matroid given in Theorem \ref{thm:setmatroid_equal_gammoid}, together with results on how to construct a generator matrix from this representation \cite{lindstrom73}. 

As we saw earlier, it is known that there exists a Singleton-optimal linear $(n,k,d,r,\delta)_i$-LRC for all parameters $(n,k,r,\delta) \in P(n,k,r,\delta)$ (cf. (\ref{eq:possible_parameters}) for a definition of $P(n,k,r,\delta)$). Further, it is also known that if $r=k$, then all Singleton-optimal linear LRCs are linear $[n,k,n-k+1]$-MDS codes. In \cite{song14} existence and nonexistence of Singleton-optimal linear $(n,k,d,r,\delta)_a$-LRCs were examined. The parameter regions for existence given in \cite{song14} were both obtained and extended in \cite{westerback15} by the construction of linear LRCs via matroid theory given in Theorem \ref{thm:matroid_to_lrc}. Hence, the results in \cite{westerback15} about nonexistence and existence of Singleton-optimal linear $(n,k,d,r,\delta)_a$-LRCs settled large portions of the parameter regions left open in $\cite{song14}$ leaving open only a minor subregion. Some improvements of the results in \cite{song14} were also given ofr $\delta=2$ in \cite{wang15} via integer programming techniques. 

For $(n,k,r,\delta) \in P(n,k,r,\delta)$ it is also very natural to ask what is the maximal value of $d$ for which there exist an $(n,k,d,r,\delta)_a$-matroid or a linear $(n,k,d,r,\delta)_a$-LRC. We denote this maximal value by $d_{max}(n,k,r,\delta)$. In \cite{westerback15} it was proven that 
$$
d_{max}(n,k,r,\delta) \geq n-k+1- \left \lceil \frac{k}{r} \right \rceil (\delta - 1)
$$ 
for linear LRCs. For matroids, this result is straightforward, as a matroid with $d=n-k+1-\left \lceil \frac{k}{r} \right \rceil (\delta - 1)$ can be constructed as a truncation of the direct sum of $\lceil \frac{n}{r+\delta-1} \rceil$ uniform matroids of size $s_i\leq r+\delta-1$ and rank $s_i-\delta+1\leq r$. As representability (over some field) is preserved under direct sums and truncation, the result follows for linear LRCs. However, with this straightforward argument, and with the bound on the field size of truncated matroids from~\cite{jurrius12}, the field size required could be as large as \[(r+\delta-1)\cdot\prod_{i=k}^{r\cdot\left\lceil\frac{n}{r+\delta-1}\right\rceil}\binom{n}{i}.\] Significant work is needed in order to bound the field size even in this special case.

This result was improved in \cite{westerback15} and further in \cite{pollanen16}. Also, the parameter region of Singleton-optimal linear $(n,k,d,r,\delta)$-LRCs was also extended in \cite{pollanen16}. The existence of Singleton-optimal linear LRCs obtained by the matroidal construction described here depends mainly on the relation between the parameters $a$ and $b$ where $a = \lceil \frac{k}{r} \rceil r - k$ and $b = \lceil \frac{n}{r+\delta-1} \rceil (r+\delta-1) - n$. Thus, we can easily get Singleton-optimal linear LRCs for all possible coding rates.

\subsection{Random Codes}
An alternative way to design $\apart$-LRCs with prescribed parameters is by exploiting the fact that independence is a generic property for $r$- and $k$-tuples of vectors over large fields. This allows us to use randomness to generate $\apart$-LRCs in a straightforward way, once the matroid structure of the code is prescribed. This is the key element in~\cite{ernvall16}. As opposed to in the gammoid construction from the last section, we will now consider the field size $q$ to be fixed but large. Indeed, a sufficiently large field will be $\F_q$ with \[q>(r\delta)^{4^rr}\binom{n+(r\delta)^{(r-1)4^r}}{k-1}.\] For given $(n,k,r,\delta)$, we will construct $\apart$-LRCs where \[d\geq n-k+1- \left \lceil \frac{k}{r} \right \rceil (\delta - 1).\] Comparing this to the generalised Singleton bound~\eqref{eq:bound_linear_nkdrd}, we notice that the codes we construct are ``almost Singleton-optimal''.

The underlying matroid will again be a truncation of \[\bigsqcup_{i=1}^{\lceil \frac{n}{r+\delta-1} \rceil} U_{s_i}^{s_i-\delta-1}.\] However, rather than first representing this direct sum, which has rank \[\sum_i (s_i-\delta +1)=n-\lceil \frac{n}{r+\delta-1} \rceil(\delta-1),\] we will immediately represent its truncation as an $n\times k$ matrix. The  random construction proceeds as follows. Divide the columns $[n]$ into locality sets $S_i$ of size $s_i$. For each $S_i$, we first generate the $r_i=s_i-\delta+1$ first columns uniformly at random from the ambient space $\F^k$. This gives us an $r_i\times k$-matrix $G_i$. After this, we draw $\delta-1$ vectors from $\F^{r_i}$, and premultiply these by $G_i$. The resulting $r_i+\delta-1=s_i$ vectors will be in the linear span of the $r_i$ first vectors, and so have rank $\leq r_i$ as a point configuration in $\F^k$. We arrange the $s_i$ vectors into a matrix $G_i '$ of rank $\leq r_i$ in $\F^{s_i\times k}$. Let $A_i$ be the event that all $r_i$-tuples of columns in $G_i$ are linearly independent. It is easy to see that, if the field size grows to infinity, the probability of $A_i$ tends to one. 

Juxtaposing the matrices $G_i '$ for $i=1,\cdots , \lceil \frac{n}{r+\delta-1} \rceil$, we obtain a generator matrix $G$ for a code of length $n$. Let $B$ be the event that $G$ has full rank. Again, assuming the field size is large enough, the probability of $B$ can be arbitrarily close to one. Now, the random matrix $G$ generates an $\apart$-LRC if all the events $A_1, \ldots, A_{\lceil \frac{n}{r+\delta-1} \rceil}, B$ simultaneously occur. A simple first moment estimate shows that, if \[q>(r\delta)^{r4^r}\binom{n+(r\delta)^{(r-1)4^r}}{k-1},\] then the probability of this is positive, so there exists an $\apart$-LRC.

\subsection{Constructing LRCs as Evaluation Codes}
As suggested in the previous sections, there are several assumptions that can be made in order to give more explicit code constructions for optimal LRCs. Next, we will follow~\cite{tamo13, tamo14} in assuming that $n$ is divisible by $r+\delta-1$ and $k$ is divisible by $r$. Then, an optimal LRC with $d=n-k-(\frac{k}{r}-1)(\delta-1)+1$ exists for any choice of $k$. We will also assume that $n=q$ is a prime power, although this assumption can easily be removed at the price of a more technical description of the code.

We will construct a Singleton-optimal code in this case as an evaluation code, generalising the construction of MDS codes as Reed-Solomon codes. The main philosophy goes back to \cite{tamo13}, but due to a technical obstacle, \cite{tamo13} still required exponential field size. This technicality was overcome by the construction in~\cite{tamo14}, which we will present next. Evaluation codes have a multitude of favourable properties, not least that the field size can often be taken to be much smaller than in na\"ive random code constructions. Moreover, the multiplicative structure used for designing evaluation codes can also be exploited when one needs to do computations with the codes in question.

Let $A$ be a subgroup of $\F_q$ of size $r+\delta-1$ and let $g=\prod_{i\in A}(x-i)$ be the polynomial of degree $r+\delta-1$ that vanishes on $A$. We will construct a storage code whose nodes are the elements of $\F_q$ and whose locality sets are the cosets of $A$. Thus, there are $\frac{n}{r+\delta-1}$ locality sets, each of size $r+\delta-1$. The codewords will be the evaluations over $\F_q$ of polynomials of a certain form.
As the rank of the code that we are designing is $k=r\cdot\frac{k}{r}$, we can write the messages as a $r\times \frac{k}{r}$ matrix \[a=\begin{pmatrix}
a_{0, 0}&\cdots & a_{r-1,0}\\
\vdots & \ddots & \vdots\\
a_{0, \frac{k}{r}-1}&\cdots & a_{r-1,\frac{k}{r}-1}
\end{pmatrix}
\]over $\F_q$. Now consider the polynomial function \[f_a=\begin{pmatrix}
1 & g(x) & g(x)^2 & \cdots & g(x)^{\frac{k}{r}-1}
\end{pmatrix}\cdot \begin{pmatrix}
a_{0, 0}&\cdots & a_{r-1,0}\\
\vdots & \ddots & \vdots\\
a_{0, \frac{k}{r}-1}&\cdots & a_{r-1,\frac{k}{r}-1}
\end{pmatrix} \cdot \begin{pmatrix}1\\ x\\ x^2\\ \vdots\\ x^{r-1}
\end{pmatrix}.
\] Consider the code \[C=\{f_a(x)\,:\,x\in \F_q,\, a\in \F_q^{r\times \frac{k}{r}}\}.\] By design, $f_a$ has degree \[\deg f_a\leq(r+\delta-1)\left(\frac{k}{r}-1\right)+r-1=k-1+(\delta-1)\left(\frac{k}{r}-1\right),\] and can therefore be computed for every point in $\F_q$ by evaluation on any $k+(\delta-1)(\frac{k}{r}-1)$ points. Therefore, the code $C$ protects against \[d-1=n-k-(\delta-1)\left(\frac{k}{r}-1\right)+1\] errors. It remains to see that it has locality $(r,\delta)$. 

To this end, note that the row vector \[\begin{pmatrix}
1 & g(x) & g(x)^2 & \cdots & g(x)^{\frac{k}{r}-1}
\end{pmatrix}\cdot \begin{pmatrix}
a_{0, 0}&\cdots & a_{r-1,0}\\
\vdots & \ddots & \vdots\\
a_{0, \frac{k}{r}-1}&\cdots & a_{r-1,\frac{k}{r}-1}
\end{pmatrix}\] of polynomials is constant over the subgroup $A\subseteq \F_q$ and thus on all of its cosets by construction of $g$. It follows that when restricted to any such coset, the function $f_a$ is a polynomial of degree $\leq r-1$, and so can be extrapolated to all points in the coset from any $r$ such evaluation points. This proves the $(r,\delta)$-locality.

As discussed, this construction depends on a collection of assumptions on the divisibility of parameters that are needed for the rather rigid algebraic structures to work. Some of these assumptions can be relaxed, using more elaborate evaluation codes, such as algebraic geometry codes over curves and surfaces \cite{barg16, barg17}. While this field of research is still very much developing, it seems that the rigidity of the algebraic machinery makes it less suitable for generalisations of the LRC concept, for example when different nodes are allowed to have different localities.


\section{Beyond Linear Storage Codes} \label{sec:beyond_linear_storage}

In this section we will introduce the notion of hierarchical codes, which are natural generalisations of locally repairable codes. After this, we will briefly describe the connection between $(n,k,d,r,\delta,t)$-LRCs and polymatroids given in $\cite{polymatroid}$.

\subsection{Hierarchical Codes}

\begin{definition}
Let $h\geq 1$ be an integer, and let $$\hpar = [(n_1,k_1,d_1, t_1), \ldots,(n_h,k_h,d_h,t_h)]$$ be a $h$-tuple of integer 4-tuples, where $k_i \geq 1$, $n_i, d_i \geq 2$, and $t_i \geq 1$ for $1 \leq i \leq h$. Then, a coordinate $x$ of a linear $\nkd=[n_0,k_0,d_0]$-LRC $C$ indexed by $E$ has \emph{$h$-level hierarchical availability} $\hpar$ if there are $t_1$ coordinate sets $X_1, \ldots, X_{t_1} \subseteq E$ such that  
$$
\begin{array}{rl}
(i) & x \in X_i \hbox{ for } i \in [t_1],\\
(ii) & i,j \in [t_1] \hcom i \neq j \Rightarrow X_i \cap X_j = \{ x  \},\\ 
(iii) & n_{X_i}\leq n_1 ,\, k_{X_i} = k_1 \hand d_{X_i} \geq d_1 \hbox{ for the punctured}\\ 
& \hbox{$[n_{X_i}, k_{X_i}, d_{X_i}]$-code $C_{X_i}$, for $i \in [t_1]$,}\\
(iv) & \hbox{for $i \in [t_1]$, $x$ has $(h-1)$-level hierarchical availa-}\\
     & \hbox{bility }  [(n_2,k_2,d_2,t_2), \ldots, (n_h,k_h,d_h,t_h)] \hbox{ in  $C_{X_i}$}.
\end{array}
$$
The code $C$ above as well as all the related subcodes $C_{X_i}$ should be non-degenerate. For consistency of the definition, we say that any symbol in a non-degenerate storage code has 0-level hierarchical availability. 
\end{definition} 

\begin{example}
Let $C$ be the code generated by the matrix $G$ in Example \ref{ex:dss_code} and let $x = 2$. Then $x$ has 2-level hierarchical availability $$\hpar = [(6,3,3,1),(3,2,2,2)]\,.$$ This follows from Example \ref{ex:nX_kX_dX-matrix} where $C_{Y_1}$ implies the $(6,3,3,1)$-availability, and the $(3,2,2,2)$-availability is implied by $C_{X_1}$ and $C_{X_2}$.
\end{example}

The most general Singleton bound for matroids with hierarchy in the case $\boldsymbol{t}=\boldsymbol{1}$ are the following given in \cite{sasidharan15,izs16}:
\[
d_i(M) \leq n_i - k_i +1-\sum_{j>i}(d_{j}-d_{j+1})\left(\left \lceil \frac{k_i}{k_{j}} \right \rceil - 1\right),
\]
where we say $d_{h+1}=1$.

\subsection{General Codes from Polymatroids}
\begin{definition}
Let $E$ be a finite set. A pair $P = (\rho,E)$ is a (finite) \emph{polymatroid} on $E$ with a \emph{set function} $\rho: 2^E \rightarrow \mathbb{R}$ if $\rho$ satisfies the following three conditions for all $X,Y \subseteq E$:
$$
\begin{array}{rl}
(R1) & \rho(\emptyset) = 0\,,\\
(R2) & X \subseteq Y \Rightarrow  \rho(X) \leq \rho(Y)\,,\\
(R3) & \rho(X) + \rho(Y) \geq \rho(X \cup Y) + \rho(X \cap Y)\,.
\end{array}
$$
\end{definition}

Note that a \emph{matroid} is a polymatroid which additionally satisfies the following two conditions for all $X \subseteq E$:
$$
\begin{array}{rl}
(R4) & \rho(X) \in \mathbb{Z}\,,\\
(R5) & \rho(X) \leq |X|\,.
\end{array}
$$ 

Using the joint entropy and a result given in \cite{fujishige78} one can associate the following polymatroid to every code.

\begin{definition}
Let $C$ be an $(n,k)$-code over some alphabet $A$ of size $s$. Then $P_C = (\rho_C, [n])$ is the polymatroid on $[n]$ with the set function $\rho_C: 2^{[n]} \rightarrow \mathbb{R}$ where
$$
\rho_C(X) = \sum_{\boldsymbol{z}_X \in C_X} \frac{|\{\boldsymbol{c} \in C: \boldsymbol{c}_X = \boldsymbol{z}_X \} |}{|C|} \log_s \left ( \frac{|C|}{|\{\boldsymbol{c} \in C: \boldsymbol{c}_X = \boldsymbol{z}_X \} |} \right )
$$
and $\rho_C(\emptyset) = 0$.
\end{definition}

We remark that for linear codes $M_C = P_C$. Using the above definition of $P_C$, one can now prove the following useful properties.

\begin{proposition}
Let $C$ be an $\ncode$ over $\alp$ with $|\alp|=s$. Then for the polymatroid $\cpoly$ and any subsets $X,Y \subseteq [n]$,
$$
\begin{array}{rl}
(i) & P_C(X) \leq |X|,\\
(ii) & |C_{X \cup Y}| > |C_X| \iff \rho_C(X \cup Y) > \rho_C(X),\\
(iii) & |C| = s^{\rho_C([n])},\\
(iv) & |C|/|\words| = s^{\rho_C([n]) - n}.
\end{array}
$$
\end{proposition}

We remark that, even though $|C| = s^{\rho_C([n])}$ for nonlinear codes and $|C_X| = s^{\rho_C(X)}$ for all $X$ for linear codes, it is not true in general that $|C_X| = s^{\rho_C(X)}$ for $X \subsetneq [n]$ for nonlinear codes. This stems from the fact that, for non-linear codes, the uniform distribution over the code does not necessarily map to the uniform distribution under coordinate projection. 

After scaling the rank function of a finite polymatroid $P = (\rho,E)$ by a constant $c$ such that $c\rho(X) \leq |X|$ for all $X \subseteq E$, we obtain a polymatroid satisfying axiom (R5). We will assume that such a scaling has been performed, so that all polymatroids satisfy axiom (R5).   

We are now ready to define a cyclic flat of a polymatroid $P = (\rho,E)$, namely $X \subseteq E$ is a \emph{cyclic flat} if 
$$
\rho(X \cup \{e\}) > \rho(X) \hbox{ for all } e \in E \setminus X \hbox{ and } \rho(X) - \rho(X \setminus \{x\}) < 1 \hbox{ for all }x \in X.
$$ 

Let $P=(\rho,E)$ be a polymatroid and $X \subseteq E$. The restriction of  $P$ to $X$ is the polymatorid $P|X = (\rho_{|X}, X)$ where $\rho_{|X}(Y) = \rho(Y)$ for $Y \subseteq X$. We can now define the distance of $P|X$ as 
$$
d(P|X) = \min \{|Y| : Y \subseteq X , \rho_{|X}(X \setminus Y) < \rho_{|X}(X)\}.
$$  

Let $\mathcal{Z}$ denote the family of cyclic flats of the polymatroid $P$. Assuming that $E \in \cflats$, we can define the parameters $n,k,d$ of $P$ via the cyclic flats and their ranks, namely 
$$
n = |E| \hcom k = \rho(E) \hand d =  \lfloor n - k + 1 - \max \{|X| - \rho(X) : X \in \mathcal{Z} \setminus E \} \rfloor.
$$ 

The definitions of $(n,k,d,r,\delta,t)_i$ and $(n,k,d,r,\delta,t)_a$-polymatroids are carried over directly from Definition \ref{def:param_matroids}. In addition, the parameters $(n,k,d,r,\delta,t)_i$ and ~$(n,k,d,r,\delta,t)_a$ of a LRC $C$ are the same as the corresponding parameters for $P_C$. Using the cyclic flats and similar methods as for matroids, Singleton-type bounds can be proven for polymatroids in general, which then imply bounds on all objects related to polymatroids, \emph{e.g.}, matroids, linear and nonlinear LRCs, and hypergraphs. This is the content of the next section.

\subsection{Singleton-type Bounds for Polymatroids and General LRCs}

It is not clear whether the Singleton-type bounds given for linear LRCs in (\ref{eq:bound_linear_nkd})--(\ref{eq:bound_linear_nkdrt}) also hold for general LRCs --- in general  the upper bound on $d$ might have to be larger. As we will describe briefly in Section \ref{sec:beyond_linear_storage}, any general LRC can be associated with a polymatroid that captures the key properties of the LRC. Using this connection we are able to define the $(n,k,d,r,\delta,t)$-parameters and information-symbol, systematic-symbol, and all-symbol locality sets for polymatroids in general. 

The class of polymatroids is much bigger than the class of the polymatroids arising from general LRCs. Hence, it is also not clear whether the Singleton-type bounds given in (\ref{eq:bound_linear_nkd})--(\ref{eq:bound_linear_nkdrt}) also hold for polymatroids in general. However, from \cite{polymatroid}, we obtain a Singleton-type bound for polymatroids in Theorem \ref{thm:polymatroid_nkdrdt} below. This theorem shows that all the Singleton-type bounds given in (\ref{eq:bound_linear_nkd})--(\ref{eq:bound_linear_nkdrt}) are polymatroid properties. Further, the polymatroid result also extends all these bounds by including all the parameters $(n,k,d,r,\delta,t)$ at the same time. 

The methods used to prove the Singleton-type bound given for polymatroids in Theorem \ref{thm:polymatroid_nkdrdt} are similar to those used for proving the Singleton-type bound for matroids in Theorem \ref{thm:bound_matroid_nkdrd}. Especially, the notion of cyclic flats is generalised to polymatroids and used as the key tool in the proof. However, some obstacles occur since we are dealing with real-valued rank functions in the case of polymatroids instead of integer-valued rank functions, which was the case for matroids. As a direct consequence of Theorem \ref{thm:polymatroid_nkdrdt}, the Singleton-type bounds given in (\ref{eq:bound_linear_nkd})--(\ref{eq:bound_linear_nkdrt}) are valid for all objects associated to polymatroids. 

\begin{theorem}[\cite{polymatroid} Singleton-type bound for polymatroids] \label{thm:polymatroid_nkdrdt}
Let $P = (\rho,E)$ be an information-set $(n,k,d,r,\delta,t)_i$-polymatroid. Then
\begin{equation} \label{eq:polymatroid_nkdrdt}
d \leq n - \lceil k \rceil + 1 - \left(  \left \lceil\frac{t(\lceil k \rceil -1)+1}{t(r-1)+1} \right \rceil - 1 \right) (\delta - 1).
\end{equation}
\end{theorem} 

Theorem \ref{thm:polymatroid_nkdrdt} is stated for information-symbol locality. This implies that the bound (\ref{eq:polymatroid_nkdrdt})  is also valid for  systematic-symbol and all-symbol locality. Hence, as a direct corollary, the bounds (\ref{eq:bound_linear_nkd})--(\ref{eq:polymatroid_nkdrdt}) hold for information-symbol, systematic-symbol, and all-symbol locality for all objects associated to polymatroids, \emph{e.g.}, entropy functions, general LRCs, hypergraphs, matroids, linear LRCs, graphs and many more. If we restrict to systematic linear codes, then the bound also holds for PIR codes \cite[Def. 4]{FVY_PIR}. The connection is not as straightforward in the nonlinear case, since the definitions of a repair group are then slightly different for LRCs (as defined here) and PIR codes, while coinciding in the linear case. 

The bound (\ref{eq:bound_linear_nkdr}), for all-symbol LRCs (as subsets of size $|B|^K$ of $B^{\alpha n}$, where $B$ is a finite set, $A = B^{\alpha}$ is the alphabet, and  $\alpha$ and $K$ are integers), follows from a result given in \cite{papailiopoulos12}. The bound (\ref{eq:bound_linear_nkdrd}), for all-symbol LRCs (as a linear subspace of $\mathbb{F}_q^{\alpha n}$ with the alphabet $A = \mathbb{F}_q^\alpha$), is given in \cite{silberstein13}. This result is slightly improved for information-symbol locality in \cite{kamath14}. The bound (\ref{eq:bound_linear_nkdrt}), for $(n,k,d,r,t)_s$-LRCs where $k$ is a positive integer, follows from a result given in \cite{rawat14}. 
The following bound for $(n,k,d,r,t)_a$-LRCs with integral $k$ was given in \cite{tamo16},
$$
d \leq n - k + 1 - \sum_{i=1}^t \left \lfloor \frac{k-1}{r^i} \right \rfloor.
$$

One parameter which has not been included above is the alphabet size. Small alphabet sizes are important in many applications because of implementation and efficiency reasons. The bound (\ref{eq:lrc_nkdr_alphabet}) below takes the alphabet size into account, but is only inductively formulated. Before stating this bound we introduce the following notation:
$$
k_{\mathrm{opt}}^{(q)}(n,d) = \max \{ k : C \hbox{ is an $(n,k,d)$-code over an alphabet of size $q$}\}.
$$
By \cite{cadambe15}, an all-symbol $(n,k,d,r)$-LRC over a finite alphabet $A$ of size $q$ satisfies
\begin{equation} \label{eq:lrc_nkdr_alphabet}
k \leq \min_{s \in \mathbb{Z}_{+}}(sr + k_{\mathrm{opt}}^{(q)}(n - s(r+1),d)).
\end{equation}
It is a hard open problem in classical coding theory to obtain a value for the parameter $k_{\mathrm{opt}}^{(s)}(n,d)$ for linear codes. This problem seems to be even harder for codes in general. However, by using other known bounds, such as the Plotkin bound or Greismer bound, it is possible to give an explicit value for $k_{\mathrm{opt}}^{(s)}(n,d)$ for some classes of parameters $(s,n,d)$. This has been done  for example  in \cite{silberstein15}. 

We remark that when considering nonlinear LRCs, some extra care has to be taken in terms of how to define the concepts associated with the LRCs. Two equivalent definitions in the linear case may differ in the nonlinear case. In this chapter, we have chosen to consider $\delta$ as a parameter for the local distance of the repair sets, \emph{i.e.}, any node in a repair set $R$ can be repaired by any other $|R| - \delta + 1$ nodes of $R$. The condition used in \cite{cadambe15, papailiopoulos12, rawat14, tamo16} is for $\delta = 2$ only assuming that a specific node in a repair set $R$ can be repaired by the rest of the nodes of $R$. It is not assumed that any node in $R$ can be repaired by the other nodes of $R$, \emph{i.e.}, that the local distance is 2. A Singleton bound using the weaker condition of guaranteeing only repair of one node in each repair set implies directly that the same upper bound on $d$ is true for the case with local distance 2. 

\section{Conclusions and Further Research}
We have shown how viewing storage codes from a matroidal perspective helps our understanding of local repairability, both for constructions and for fundamental bounds. However, many central problems about linear LRCs boil down to notoriously hard representability problems in matroid theory.

A famous conjecture, with several consequences for many mathematical objects, is the so called \emph{MDS-conjecture}. This conjecture states that, for a given finite field $\F_q$ and a given $k$, every $[n,k,d]$-MDS code over $\F_q$ has $n\leq q+1$, unless in some special cases. Currently, the conjecture is known to hold only if $q$ is a prime~\cite{BallMDS}. Linear Singleton-optimal LRCs may be seen as a generalisation of linear MDS codes. An interesting problem would therefore be to consider an upper bound on $n$ for linear Singleton-optimal LRCs over a certain field size $q$ with fixed parameters $(k,r,\delta,t)$. In this setting, a sufficiently good upper bound on $n$ would be a good result.

Instead of fixing the Singleton-optimality and trying to optimise the field size, we could also fix the field $\F_q$, and try to optimise the locality parameters. This would give us bounds on the form \[d \leq n - k + 1 - \left( \left \lceil \frac{k}{r} \right \rceil - 1 \right) (\delta - 1) - p(q,n,k,r,\delta),\] where the dependence on the field size $q$ is isolated to a ``penalty'' term $p(q,n,k,r,\delta)$. Partial results in this direction are given by the Cadambe-Mazumdar bound~\cite{cadambe15}, and LRC versions of the Griesmer and Plotkin bounds~\cite{silberstein15}.  However, the optimality of these bounds is only known for certain ranges of parameters. Further research in this direction is definitely needed, but seems to lead away from the most obvious uses of matroid theory.

Finally, it would be interesting to characterise all Singleton-optimal LRCs up to matroid isomorphism. The constructions discussed in this paper appear to be rather rigid, and unique up to shifting a few ``slack'' elements between different locality sets. However, it appears to be difficult to prove that all Singleton-optimal matroids must have this form. Once a complete characterisation of Singleton-optimal matroids has been obtained, this could also be taken as a starting point for possibly finding Singleton-optimal nonlinear codes in the parameter regimes where no Singleton-optimal linear codes exist.

%

 \bibliographystyle{spmpsci}
 \bibliography{references2}
 
 \section*{Appendix: More about Matroid Theory}
 A matroid realisation of an $\F$-linear matroid $M$ has two geometric interpretations. Firstly, we may think of a matrix representing $M$ as a collection of $n$ column vectors in $\F^k$. As the matroid structure is invariant under row operations, or in other words under change of basis in $\F^k$, we tend to think of $M$ as a configuration of $n$ points in abstract projective $k$-space. 

The second interpretation comes from studying the row space of the matrix, as an embedding of $\F^k$ into $\F^n$. Row operations correspond to a change of basis in $\F^k$, and hence every matroid representation can be thought of as a k-dimensional subspace of $\F^n$. In other words, a matroid representation is a point in the Grassmannian $\gr(n,k; \F)$, and $\gr(n,k; \F)$ has a stratification as a union of realisation spaces $R(M)$, where $M$ ranges over all $\F$-representable matroids of size $n$  and rank $k$. This perspective allows a matroidal perspective also on the subspace codes discussed in Chapter 1--4, where the codewords themselves are matroid representations. However, so far this perspective has not brought any new insights to the topic.

Another instance where matroids appear naturally in mathematics is graph theory. Let $\Gamma$ be a finite graph with edge set $E$. We obtain a matroid $M_\Gamma=(\indeps, E)$, where $I\subseteq E$ is independent if the subgraph $\Gamma_I\subseteq\Gamma$ induced on $I\subseteq E$ is a forest, \emph{i.e.}, has no cycles. A matroid that is isomorphic to $M_\Gamma$ for some graph $\Gamma$ is said to be a \emph{graphical} matroid.

\begin{example} \label{ex:indexed_graph}
The matrix $G$ and the graph $\Gamma$ given below generate the same matroid, regardless of the field over which $G$ is defined.

\[
G=
\begin{tabular}{ |c|c|c|c|c|c|c| }
\multicolumn{1}{c}{1}&
\multicolumn{1}{c}{2}&
\multicolumn{1}{c}{3}&
\multicolumn{1}{c}{4}&
\multicolumn{1}{c}{5}&
\multicolumn{1}{c}{6}&
\multicolumn{1}{c}{7}\\
\hline
1&0&0&0&0&1&1\\
\hline
0&1&0&0&0&1&1\\
\hline
0&0&1&0&-1&-1&0\\
\hline
0&0&0&1&1&0&-1\\
\hline
\end{tabular}\ ,
\hspace{0.3cm}
\begin{tikzpicture} 
    \node (1) [circle, draw, fill=black!50, inner sep=0pt, minimum width=4pt] {};
    \node (2) [circle, draw, fill=black!50, inner sep=0pt, minimum width=4pt] [above left of=1] {};
    \node (3) [circle, draw, fill=black!50, inner sep=0pt, minimum width=4pt][above right of=1] {}; 
    \node (4) [circle, draw, fill=black!50, inner sep=0pt, minimum width=4pt][above  of=1] {}; 
    \node (5) [circle, draw, fill=black!50, inner sep=0pt, minimum width=4pt][above  of=4] {};
    \node (6) [left of=2] {$\Gamma=$};
    \path[-]
    (1) edge node[left]{4} (2)
    (1) edge node[above right]{5} (2)
    (1) edge node[right]{7} (3)
    (4) edge (2)
    (3) edge (4)
    (2) edge node[left]{1} (5)
    (3) edge node[right]{2} (5)
    (1) edge node[above left=0.2cm]{3} (4)
    (1) edge node[above right=0.2cm]{6} (4);    
\end{tikzpicture}
\]
Some examples of independent sets in $G$ and $\Gamma$ are $\{3,4,6\}, \{1,2,3,5\}, \{2,3,4,6\}$. The set $X = \{5,6,7\}$ is dependent in $M_\Gamma$ as these edges form a cycle, and it is dependent in $M_G$ as the submatrix $$
G(X)=
\begin{tabular}{|c|c|c|}
\multicolumn{1}{c}{5}&
\multicolumn{1}{c}{6}&
\multicolumn{1}{c}{7}\\
\hline
0&1&1\\
\hline
0&1&1\\
\hline
-1&-1&0\\
\hline
1&0&-1\\
\hline
\end{tabular}
$$ has linearly dependent columns.
\end{example}

Indeed, graphical matroids are representable over any field $\F$. To see this, for a graph $\Gamma$ with edge set $E$, we will construct a matrix $G(\Gamma)$ over $\F$ with column set $E$ as follows. Choose an arbitrary spanning forest $T\subseteq E$ in $\Gamma$, and index the rows of $G(\Gamma)$ by $T$. Thus $G(\Gamma)$ is a $T\times E$-matrix. Choose an arbitrary orientation for each edge in the graph. For $e\in T\subseteq E$ and $uv\in E$, the entry in position $(e,\{uv\})$ is $1$ (respectively $-1$) if $e$ is traversed forward (respectively backward) in the unique path from $u$ to $v$ in the spanning forest $T$. In particular, the submatrix $G(\Gamma)(T)$ is an identity matrix. It is straightforward to check that the independent sets in $G(\Gamma)$ are exactly the noncyclic sets in $\Gamma$. 

\begin{example}
The matrix $G$ in Example \ref{ex:indexed_graph} is $G(\Gamma)$ where $\Gamma$ is the graph in the same example, and the spanning forest $T$ is chosen to be $\{1,2,3,4\}$.
\end{example}

The restriction to $X\subseteq E$ of a graphical matroid $M_\Gamma$ is obtained by the subgraph of $\Gamma$ containing precisely the edges in $X$.

\begin{wrapfigure}{r}{0.36\textwidth}
\begin{center}
\includegraphics[scale=.25]{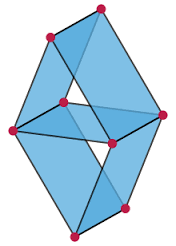}
\end{center}
\caption{The Vamos matroid of size $8$ and rank $4$, which is not algebraically representable.}
\label{fig:vamos}       
\end{wrapfigure}

A third example of matroids occurring naturally in mathematics are \emph{algebraic matroids}~\cite{algebraic}. These are associated to field extensions $\F:K$ together with a finite point sets $E\subseteq K$, where the independent sets are those $I\subseteq E$ that are algebraically independent over $\F$. In particular, elements that are algebraic over $\F$ have rank zero, and in general $\rho(I)$ is the transcendence degree of the field extension $\F(I):\F$.

It is rather easy to see that every $\F$-linear matroid is also algebraic over $\F$. Indeed, let $X_1,\cdots , X_k$ be indeterminates, and let $$g:\F^k\to \F(X_1,\cdots , X_k)$$ be given by $\ee_i\mapsto X_i$ for $i=1,\cdots k$. Then $J\subseteq E$ is linearly independent over $\F$ if and only if $\{g(j):j\in J\}$ is algebraically independent over $\F$. 
Over fields of characteristic zero the converse also holds, so that all algebraic matroids have a linear representation. However, in positive characteristic there exist algebraic matroids that are not linearly representable. For example, the non-Pappus matroid of Example~\ref{ex:pappus} is algebraically representable over $\F_4$, although it is not linearly representable over any field~\cite{lindstrom88}. The smallest example of a matroid that is not algebraic over any field is the Vamos matroid, in Figure~\ref{fig:vamos}~\cite{ingleton75}.

\begin{definition}
The \emph{dual} of $\rmatroid$ is $M^*=(\rho^*,E)$, where $$\rho^*(X)=|X|+\rho(E\setminus X) -\rho(E).$$ 
\end{definition}
The definition of the dual matroid lies in the heart of matroid theory, and has profound interpretations. In geometric terms, let $M$ be represented by a $k$-dimensional subspace $V$ of $\F^n$. Then, the matroid dual $M^*$ is represented by the orthogonal complement $V^\perp\subseteq \F^n$. Surprisingly and seemingly unrelatedly, if $\Gamma$ is a planar graph and $M=M_\Gamma$ is a graphical matroid, then  $M^*=M_{\bar{\Gamma}}$, where $\bar{\Gamma}$ is the planar dual of $\Gamma$. Moreover, the dual $M_\Gamma^*$ of a graphical matroid is graphical if and only if $\Gamma$ is planar.

\begin{definition}
The \emph{contraction} of $X\subseteq E$ in the matroid $\rmatroid$ is $M/X=(\rho',M\setminus X)$, where $\rho'(Y)=\rho(Y\cup X)-\rho(X)$.
\end{definition}
Contraction is the dual operation of deletion, in the sense that $M/X=(M^*_{|E\setminus X})^*$. The terminology comes from graphical matroids, where contraction of the edge $e\in E$ corresponds to deleting $e$ and identifying its endpoints in the graph. Notice that it follows directly from submodularity of the rank function that $\rho_{M/X}(Y)\leq \rho_{M|_{E\setminus X}}(Y)$ for every $Y\subseteq E\setminus X$.
In terms of subspace representations, contraction of $e\in E$ corresponds to intersecting the subspace that represents $M$ with the hyperplane $\{x_e=0\}$.

As matroids are used as an abstraction for linear codes, it would be desirable to have a way to go back from matroids to codes, namely to determine whether a given matroid is representable, and when it is, to find such a representation. Unfortunately, there is no simple criterion to determine representability~\cite{vamos78, mayhew14}. However, there are a plethora of sufficient criteria to prove nonrepresentability, both over a given field and over fields in general. In recent years, these methods have been used to prove two long-standing conjectures, that we will discuss in Sections~\ref{subsec:rota} and~\ref{subsec:nelson} respectively.

\subsection{Rota's Conjecture}
\label{subsec:rota}

While there is no simple criterion to determine linear representability, the situation is much more promising if we consider representations over a fixed field. It has been known since 1958, that there is a simple criterion for when a matroid is binary representable. 

\begin{theorem}[\cite{tutte58}]
Let $\rmatroid$ be a matroid. The following two conditions are equivalent.
\begin{enumerate}
\item $M$ is linearly representable over $\F_2$.
\item There are no sets $X\subseteq Y\subseteq E$ such that $M|Y/X$ is isomorphic to the uniform matroid $U_4^2$.
\end{enumerate}
\end{theorem}

In essence, this means
that the only obstruction that needs to be overcome in order to be representable over the binary
alphabet, is that no more than three nonzero points can fit in the same plane. For further reference, we say that a \emph{minor} of the matroid $\rmatroid$ is a matroid of the form $M|Y/X$, for $X\subseteq Y\subseteq E$. Clearly, if $M$ is representable over $\F$, then so is all its minors. Let $L(\F)$ be the class of matroids that are not representable over $\F$, but such that all of their minors are $\F$-representable. Then the class of $\F$-representable matroids can be written as the class of matroids that does not contain any matroid from $L(\F)$ as a minor. Gian-Carlo Rota conjectured in 1970 that $L(\F)$ is a finite set for all finite fields $\F$. A proof of this conjecture was announced by Geelen, Gerards and Whittle in 2014, but the details of the proof still remain to written up~\cite{whittle14}. 

\begin{theorem}\label{thm:rota}
For any finite field $\F$, there is a finite set $L(\F)$ of matroids such that any matroid $M$ is representable if and only if it contains no element from $L(\F)$ as a minor.\end{theorem}

Since the 1970's,
it has been known that a matroid is representable over $\F_3$ if and only if it avoids the uniform
matroids $U_5^2$, $U_5^3$, the Fano plane $P^2(\F_2)$, and its dual $P^2(\F_2)^*$ as minors. The list $L(\F_4)$ has seven elements, and was given explicitly in 2000. For larger fields, the explicit list is not known, and there is little hope to even find useful bounds on its size.

\subsection{Most Matroids are Nonrepresentable}\label{subsec:nelson}
For a fixed finite field $\F$, it follows rather immediately from the minor-avoiding description in the last section that the fraction of $n$-symbol matroids that is $\F$-representable goes to zero as $n\to \infty$. It has long been a folklore conjecture that this is true even when representations over arbitrary fields are allowed. However, it was only in 2016 that a verifiable proof of this claim was announced~\cite{nelson16}. 
\begin{theorem}\label{thm:almostall}
$$\lim_{n \rightarrow \infty} \frac{\# \hbox{linear matroids on $n$ elements}}{\# \hbox{matroids on $n$ elements}} = 0.$$
\end{theorem}

The proof is via estimates of the denominator and enumerator of the expression in~\ref{thm:almostall} separately. Indeed, it is shown in~\cite{knuth74} that the number of matroids on $n$ nodes is at least $\Omega(2^{(2-\varepsilon)^n})$ for every $\epsilon>0$. The proof of Theorem~\ref{thm:almostall} thus boiled down to proving that the number of representable matroids is $O(2^{n^3})$. This is in turn achieved by bounding the number of so called \emph{zero-patterns} of polynomials.

\subsection{Gammoid Construction of Singleton-Optimal LRCs}
For completeness, we end this appendix with a theorem that explicitly presents the matroids constructed in Theorem \ref{thm:matroid_construction} as gammoids. As discussed in Section4.2, this proves the existence of Singleton-optimal linear LRCs whenever a set system satisfying~\eqref{eq:matroid_construction_conditions} exists.
\begin{theorem} [\cite{westerback15}, $\setmatroid$-matroids are gammoids]\label{thm:setmatroid_equal_gammoid}
Let $M(F_1,\ldots,F_m;\rho)$ be a matroid given by Theorem \ref{thm:matroid_construction} and define $s: E \rightarrow 2^{[m]}$ where $s(x) = \{i \in [m] : x \in F_i\}$. Then $\setmatroid$ is equal to the gammoid $\gammoid$, where $\Gamma = (V,D)$ is the directed graph with
$$
\begin{array}{rl}
(i) & V = E \cup H \cup T \hbox{ where } E,H,T \hbox{ are pairwise disjoint},\\
(ii) & T = [k],\\
(iii) & H \hbox{ equals the union of the pairwise disjoint sets}\\
& \hbox{$H_1, \ldots ,H_m, H_{\geq 2}$, where}\\
& \begin{array}{l}
|H_i| = \rho(F_i) - |\{x \in F_i : |s(x)| \geq 2\}| \hbox{ for } i \in [m],\\
H_{\geq 2} = \{h_y : y \in E \hcom |s(y)| \geq 2 \},\\ 
\end{array}\\
(iv) & D = D_1 \cup D_2 \cup D_3 \hbox{, where}\\
& \begin{array}{l}
D_1 = \bigcup_{i \in [m]}\{(\overrightarrow{x,y}) : x \in E \hcom s(x) = \{i\} \hcom y \in H_i\},\\
D_2 = \{(\overrightarrow{x,h_y}) : x \in E \hcom h_y \in H_{\geq 2} \hcom s(x) \subseteq s(y) \},\\
D_3 = \{(\overrightarrow{x,y}) : x \in H, y \in T\}.\\
\end{array}
\end{array}
$$
\end{theorem}

\end{document}